# Stellar Companions to Stars with Planets

J. Patience[1], R. J. White[2], A. M. Ghez[3], C. McCabe, I. S. McLean, J. E. Larkin, L. Prato, Sungsoo S. Kim

Department of Physics and Astronomy, University of California at Los Angeles, Los Angeles, CA 90095-1562

J. P. Lloyd, M. C. Liu[4], J. R. Graham

Department of Astronomy, University of California at Berkeley, Berkeley, CA 94720

B. A. Macintosh, D. T. Gavel, C. E. Max, B. J. Bauman, S. S. Olivier

Institute of Geophysics and Planetary Physics, Lawrence Livermore National Lab, L-413, 7000 East Ave., Livermore, CA 94550

P. Wizinowich & D. S. Acton

W. M. Keck Observatory, CARA, 65-1120 Mamalahoa Highway, Kamuela, HI 96743

[1] present address: IGPP, LLNL L-413, 7000 East Ave., Livermore, CA 94550

[2] present address: McDonald Observatory, R. L. M. Hall 15.308, Austin, TX 78712-1083

[3] also: IGPP, University of California at Los Angeles, Los Angeles, CA 90095

[4] present address: currently Beatrice Watson Parrent Fellow, Institute for Astronomy, 2680 Woodlawn Dr., University of Hawaii, Honolulu, HI 96822

Abstract

A combination of high-resolution and wide-field imaging reveals two binary stars and one triple star system among the sample of the first 11 stars with planets detected by radial velocity variations. High resolution speckle or adaptive optics (AO) data probe subarcsecond scales down to the diffraction limit of the Keck 10-m or the Lick 3-m, and direct images or AO images are sensitive to a wider field, extending to 10" or 38", depending upon the camera. One of the binary systems – HD 114762 – was not previously known to be a spatially resolved multiple system; additional data taken with the combination of Keck adaptive optics and NIRSPEC are used to characterize the new companion. The second binary system – Tau Boo – was a known multiple with two conflicting orbital solutions; the current data will help constrain the discrepant estimates of periastron time and separation. Another target – 16 Cyg B – was a known common proper motion binary, but the current data resolve a new third component, close to the wide companion 16 Cyg A. Both the HD 114762 and 16 Cyg B systems harbor planets in eccentric orbits, while the Tau Boo binary contains an extremely close planet in a tidally-circularized orbit. Although the sample is currently small, the proportion of binary systems is comparable to that measured in the field over a similar separation range. Incorporating the null result from another companion search project lowers the overall fraction of planets in binary systems, but the detections in our survey reveal that planets can form in binaries separated by less than 50 AU.

1. Introduction

Ongoing radial-velocity discoveries of extrasolar planets continue to increase the number of solar-type stars with known planets, and the most recent accounting lists more than 90 confirmed systems *(http://www.obspm.fr/encycl/catalog.html)*. With the growing number of systems known, it is possible to begin to investigate systematically the environments of stars with planets. Given the high frequency of binary stars -- ~100% (e.g. Ghez et al. 1993; Simon et al. 1995) of T Tauri stars in nearby star forming regions and ~58% (Duquennoy & Mayor 1991) of G-dwarfs in the solar neighborhood – it is likely that companion stars influence both star and planet formation. Studies comparing the millimeter and sub-millimeter emission from T Tauri single and binary stars find decreased emission from binaries with separations <50-100 AU, suggesting that close companion stars may significantly disrupt circumstellar disks and lower the potential to form planets (e.g. Jensen et al. 1996). Although the total disk mass in the binaries may be depleted relative to singles, resolved multiwavelength *Hubble Space Telescope (HST)* observations of T Tauri binaries separated by 10-100 AU demonstrate that the components of many of these systems are able to support limited circumstellar disks (White & Ghez 2001). The known extrasolar planetary systems provide a unique dataset to directly test whether or not planets – specifically giant planets with semimajor axes less than a few AU – can form in binary systems and how companion stars affect planetary properties. Companion searches are particularly important for the systems with radial velocity residuals suggestive of longer period companions and those in eccentric orbits which may have been produced by perturbations from a companion star or massive planet (e.g. Mazeh et al. 1997; Artymowicz 1998).

By combining speckle, adaptive optics (AO), and direct images, this study is designed to characterize the binary status of the stars with known planets. The bright target magnitudes (V=4.1-10.2 mag) are a practical consideration for this project, since the speckle technique requires very short integration times and adaptive optics systems provide better wavefront correction for brighter targets. Although the magnitude limit of the Lick and Keck AO systems currently extends to *V~13* mag (Gavel et al. 1999), these bright stars presented an ideal initial target set for the AO systems as they were being developed. The results demonstrate the success of new techniques and instrumentation: the AO-optimized infrared camera IRCAL (Lloyd et al. 2000) in combination with the Lick AO system built by Lawrence Livermore National Lab (LLNL) (Bauman et al. 1999; Gavel et al. 1999) and the cryogenic infrared spectrograph NIRSPEC (McLean et al. 1998) with the Keck AO system (Wizinowich et al. 2000).

2. Sample

The sample for this project consists of the first 11 stars with planets detected by radial velocity variations. The observed targets are a representative subset of the known extrasolar planetary systems, spanning the entire range of orbital semi-major axes, covering a variety of companion *msini* values, and including both circular and eccentric systems. Table 1 summarizes the basic properties of the target stars and their planetary mass companions. For the target stars, the coordinates, magnitudes, and distances are taken from SIMBAD, while the spectral types are from the references listed in the fifth column. For the planetary companions, the minimum mass, semi-major axis, eccentricity, and notes about radial velocity residuals are those reported in the discovery papers given in the last column. An important selection effect for the radial velocity surveys is the exclusion of targets with known companions with separations small enough to contaminate the primary spectrum (<2" Marcy et al. 2000).

3. Observations

3.1 Imaging

Both high resolution and wide field images were obtained for each target. Using the highest resolution techniques, the speckle and adaptive optics observations are sensitive to subarcsecond companions; each target was observed with at least one of the following high resolution systems: Keck speckle, Lick AO, or Lick speckle. For the speckle observations, several hundred short exposures (~0.1s) of the target star and a point source were recorded and a Fourier analysis of these snapshots produces diffraction-limited images of the object (Labeyrie 1970, Lohmann et al. 1983). The smallest detectable separation for the targets observed with Keck speckle is 0".023, set by half the diffraction-limit of a 10m telescope at a wavelength of 2.2 μm.; the corresponding limit for the Lick speckle data is 0".076.

Additional high resolution data were taken with the Lick AO system. The LLNL AO system has 61 actively-controlled degrees of freedom and operating on these bright stars, it produced images with Strehl ratios in the 0.3-0.6 range, depending upon seeing, and diffraction-limited cores with FWHM 0".15 at $K_S(\lambda_c=2.1\mu m)$. Unsaturated images were recorded through a narrow-band filter for sensitivity to subarcsecond companions, and deep saturated broadband images were obtained to search for fainter companions at larger separations. The IRCAL camera includes a focal-plane occulting finger which was placed over the target star in the deep images.

A comparison PSF star to characterize the system performance and to distinguish PSF artifacts was also observed with each AO target. The PSF stars were selected to have similar declination as the target and a right ascension offset by ~30min so that the airmass at the time of observation matched that of the target star. Whenever possible, the PSF star had a very similar magnitude and color. Because the targets are such bright stars, some PSF stars were up to 1 magnitude fainter. In all cases, however, the wavefront sensor signal was very high and the AO system was run at the highest rate (500 Hz) and with the same gain for target and PSF.

Direct images at Keck also provided wide field data; these images were taken through narrow-band filters in an attempt to avoid saturation. One system – 16 Cyg B – was also observed through broadband filters and a photometric standard was used for calibration. Two targets – 55 Cnc and 47 UMa – lack wider field images from the current data, however, coronagraphic searches for dust around 55 Cnc (Trilling & Brown 1998, Schneider et al. 2000) would have revealed any stellar companion to that extrasolar planet star . Details of the different instruments are listed in Table 2.

Because of the proximity of these stars (D=4.70-40.6 pc), the projected separations covered by the data both overlap the binary separation distribution peak (~30 AU, Duquennoy & Mayor 1991) and encompass the size associated with circumstellar disks (~100 AU; Shu et al. 1987); companions in this separation range are believed to influence the evolution of protoplanetary disks. For the median distance of 15pc, the angular separation range covered for each star -- 0".15 to 10".0 -- makes the data sensitive to stellar companions from 2.25 AU to 150 AU.

For the targets with candidate companions, additional measurements were made to discriminate between physically associated companions and background objects. Multiwavelength images were obtained to determine the color and relative magnitude of possible companions, and a second epoch image was taken to confirm that the companion exhibited a common proper motion. At Keck, images through either *J ($\lambda_c$=1.25μm)* and *K ($\lambda_c$=2.2μm)* or the narrow *Brγ ($\lambda_c$=2.165μm)* and *OII* filters *($\lambda_c$=1.236μm)* were taken. At Lick, *J, H ($\lambda_c$=1.65μm)*, and *$K_S$ ($\lambda_c$=2.1μm)* images of the binaries were taken with a neutral density filter with an attenuation factor of 20. Table 3 lists the imaging observations made for each target.

3.2 Spectroscopy

Infrared spectroscopy of one object, the companion to HD114762, was obtained by McLean, Larkin, and Prato during commissioning of the NIRSPEC near-infrared spectrometer with the Keck Adaptive Optics system on June 5, 2000. NIRSPEC was used in its low-resolution mode ($R=\lambda/\Delta\lambda\sim2500$ for a 2-pixel slit) to obtain spectra in the *J, H*, and *K* bands of the companion. A detailed description of NIRSPEC is given elsewhere (McLean et al. 1998). The projected slit width corresponded to 0".036 and AO performance gave images with FWHM of less than 0".050. For each waveband, a pair of 300s exposures was obtained with the object located at two different positions along the slit. To correct for absorption due to the Earth's atmosphere, an A1 star (HD 119055) was observed at the same airmass. Immediately after each observation, arc lamp and white light sources were observed for wavelength and flat-field calibrations. Reduction of the NIRSPEC data followed the same steps as outlined in McLean et al. 2000.

## 4. Results

### 4.1 Singles

For the targets without a spatially resolved companion, the magnitude difference detection limits were determined at several angular separations. The resulting limits for each program star are given in Appendix A and a summary of the survey sensitivity is given in Figure 1. The procedure for the speckle data involves fitting cosine patterns to the calibrated power spectra as described in Ghez et al. (1993 and 1997). For this study, the speckle data are typically most sensitive at the closest separations followed by adaptive optics and direct imaging.

The detection limits for the images are based on the variation in the background at each radius from the target star; a PSF subtraction was performed to aid in identifying companions in the AO images, but the detection limits are measured from the unsubtracted target star image. To calculate the limits for AO or direct imaging, each pixel was replaced by the median of a FWHM-sized box centered on the pixel and multiplied by the number of pixels in the box. The standard deviation of the median pixels at a given radius corresponds to a 1 $\sigma$ detection. Before calculating the standard deviation, the pixels in the brightest NIRC diffraction spike, in the IRCAL occulting finger, or associated with companions were masked out. Additionally, the saturated IRCAL images were subtracted by the median-filtered images to remove scattered light. Based on this detection method, the faintest companion, HD 114762 B, is a 7 $\sigma$ detection in the NIRC images and a 13 $\sigma$ detection in the IRCAL images; 5 $\sigma$ is used as the detection threshold. The 5 $\sigma$ threshold was set by scaling down the flux of HD 114762 B by successive numbers of the standard deviation at its radius. Though the companion was marginally distinct at 3 $\sigma$, by 5 $\sigma$ it was clearly detected by visual inspection.

For a given separation one of the techniques (speckle/AO/direct imaging) gives the best detection limit for each star. Using the most sensitive detection limit for each star, the median sensitivity at a series of separations is shown in Figure 1. The error bars represent one standard deviation of the best limits. The individual target sensitivity limits used to construct Figure 1 are given in Appendix A. Only stars observed with Keck contribute to the separations less than 0".15. At the minimum separation considered for the entire survey, 0".15, the observations have a detection limit of $\Delta K \sim 4$ mag which corresponds to a mass ratio limit of $mass_{sec}/mass_{prim} \sim 0.25$. Beginning at a separation of 1".5, the median limit of ~7 mag corresponds to mass ratios which reach the bottom of the main sequence for the solar-mass targets.

Although the radial velocity searches should detect most stellar mass companions at the closer separations of this project, there remain some inclination angles to which the radial velocity observations are insensitive. For the wider separations covered by this survey, the corresponding periods are large enough that stellar companions would not be expected to produce a significant residual over the time period the high precision radial velocity studies have been conducted.

4.2 Binaries

Imaging

These observations revealed physical companions to HD 114762, Tau Boo and 16 Cyg A; an unassociated background object was also imaged near 16 Cyg B. For the companions to HD 114762, Tau Boo, and 16 Cyg, the separations, position angles, and magnitude differences of the systems were determined from the imaging data. For the Keck direct imaging data, the IRAF package *apphot* was used to perform photometry with small apertures (~0".6) and to measure the separations and position angles listed in Table 4. The values represent the average of the 5 individual images at each wavelength and the uncertainty is the standard deviation of the measurements combined with an additional 1% to account for NIRC and IRCAL pixel scale uncertainties. Because the targets were dithered to several positions on the array and the measured separations are small compared to the size of the entire field, the uncertainties should include some of the variations in the pixel scale. Based on detailed ray-tracing or comparison with known multiple stars, the variation in pixel scale across the field-of-view ranges from at most 2% for NIRC (appendix of Ghez et al. 1998) to 10% for LIRCII (Trapezium measurements from Macintosh); these uncertainties do not impact the assessment of companion association, as they are very small compared to the expected motion of a background object given the large proper motions of the targets. The J and K magnitude differences reported for 16 Cyg B assume J=5.04 mag and K=4.65 mag for the saturated primary, while the magnitude of the companion is based on comparison with a photometric standard. Due to saturation in the broadband images, the separation and position angle for this target are only determined with the narrow-band image. For the Lick AO data a similar procedure was followed using IDL routines. The measured angular separations, position angles, and magnitude differences are listed in Table 4.

The HD 114762 and 16 Cyg A companions are new results, while the Tau Boo companion is the catalogued binary system ADS 9025. The J and K magnitude differences were enough to rule out the faint, wide companion to 16 Cyg B, however, proper motion measurements were required to assess the association of the remaining double stars. With the high proper motions of these nearby target stars, typically ~0".5/yr, the two-year baseline between imaging observations would show a substantial difference in positions for background objects. Based on the photometry and proper motion, the companions to HD 114762, Tau Boo, and 16 Cyg A are physically associated. Figure 2 shows two epochs of observations on the newly resolved companion to HD 114762.

Spectroscopy

In the NIRSPEC J-band spectrum (1.135-1.362 µm) of the companion to HD114762 (see Figure 3) the dominant lines are transitions of neutral potassium (KI) at 1.1690, 1.1770, 1.2432, and 1.2522 µm. Lines of neutral metals such as Fe I and Al I are also prominent and there is a moderately strong band of FeH at 1.194 µm. In addition, a strong absorption band due to steam ($H_2O$) occurs at about 1.33 µm.

The strength of the steam band as measured by the ratio of the fluxes at 1.34 and 1.29 µm has been found to be a good indicator of spectral type from M5 through the L-dwarf sequence (McLean et al. 2000, Reid et al. 2001). In the companion to HD 114762, this ratio (~0.6) would imply a late M or early L spectral type. However, this result is inconsistent with the strength of the FeH bands, which should be much stronger if the spectral type is late M/early L. There are other inconsistencies too. The neutral metal lines (Fe I and Al I) in the companion to HD 114762 are unusually strong, which is more characteristic of an earlier spectral type. The potassium line strengths are consistent with a broader range of late M types. By comparing the spectrum in Figure 3 with NIRSPEC spectra of GL 406 (M5.5), LHS 2065 (M9) and 2MASS0746 (L0.5), the most likely spectral type is M6 rather than M9/L0. In summary, the strength of the steam band is consistent with NIRSPEC observations of early L-dwawrfs, however, the refractory metals are usually missing at such lower temperatures and the FeH band is much stronger. Without the steam band, the spectrum is similar to that of a mid-M star such as GL406. This is a curious result which requires further observations.

5. Discussion

5.1 Individual Binaries

5.1.1 Tau Boo

Based on CSHELL data of Tau Boo A, Wiedemann et al. (2000) have suggested that an observed methane signature in the spectrum could be due to a substellar companion, but this interpretation depends upon the amount of contamination from the secondary through the slit. The basic parameters of the orbit are uncertain, with two conflicting solutions. The Keck direct image was taken only 8 months after the CSHELL data and the measured separation, 2".87, and position angle, 30.8 deg., of the Keck data suggest that the companion would not contribute significantly to the spectrum of the primary. The current data do not suggest that the predicted periastron separations occurred earlier than expected, as proposed by Wiedemann et al. Continued monitoring will empirically determine the periastron distance. The orbit calculated by Popvic & Palovic (1996) predicts separation of 2".52 and position angle of 38.57 deg. at the time of the observation, while the Hale (1994) orbit predicts a similar separation of 2".68, but a smaller position angle of 31.2 deg. The observed separation is larger than either calculation, but the position angle is consistent with the Hale solution.

Combining the orbital solutions for Tau Boo with the results from numerical simulations of disks in binary systems, the maximum size of the circumprimary and circumsecondary disks originally present in the system can be estimated. Based on simulations of the interactions between binary stars and disks (Artymowicz & Lubow 1994, Clarke & Pringle 1993), the circumstellar disks around each component of a binary system should be truncated to a fraction of the semi-major axis depending upon the mass ratio and eccentricity of the binary orbit. With an F7 primary and early-M secondary, the Tau Boo system is best represented by the simulations involving $\mu=0.30$ in the Artymowicz and Lubow study (1994). Because of the different predictions for semi-major axis and eccentricity, the two calculations of the Tau Boo orbital parameters have somewhat different implications for the circumstellar disks. The Popvic & Palovic orbit semi-major axis is 6".3001 or 98.3 AU and the eccentricity is 0.4189. Over the range of disk gas Reynolds numbers considered in the simulations, the approximate edge of the circumprimary disk for Tau Boo is 18-29 AU and the circumsecondary disk cutoff is about 14-19 AU. Disks of this size are compatible with giant planet formation initiated by the accumulation of icy planetesimals outside the ice condensation radius (Bodenheimer & Pollack 1986), Although the Tau Boo secondary is a cooler M star, calculations of the location of the ice condensation radius around G to M stars (Boss 1995) show only a weak dependence on mass and range from 6 AU for a 1.0 Msun star (Tau Boo A ~1.3 Msun) to 4.5 AU for a 0.1 Msun star (Tau Boo B ~0.4Msun). The Hale orbit semi-major axis is 14".39 or 225 AU and the eccentricity is

0.91, giving a periastron distance of 20 AU. The Artymowicz and Lubow simulations do not extend to such high eccentricities, but simulations of disks around stars that experience a stellar fly-by (Clarke & Pringle 1993) are appropriate to this orbit (although the simulations assume similar masses for the two stars). The coplanar, prograde case considered in the Clarke and Pringle study (1993) is most similar to an eccentric orbit and results in a primary disk that loses some material from radii as small as 0.35 times periastron, 7 AU for Tau Boo. The majority of the disk outside 0.5 times periastron, 10 AU for Tau Boo, is stripped. The limited disk material may have prevented the formation of a multiple planet system such as those detected in 4 of the 11 stars observed (Fischer et al. 2002, Marcy et al. 2002, Marcy et al. 2001, Butler et al. 1999). If enough disk material outside the ice condensation radius was stripped by the repeated encounters of the two stars (Hale orbital period 2000 yrs), then the planet formation that occurred around Tau Boo may have required a different mechanism such as gravitational instabilities in the disk (Cameron 1978).

5.1.2 HD 114762

Among the stars with radial velocity detected planets, HD 114762 is atypical, as it has a low metallicity of only [Fe/H]=-0.60 (Gonzalez 1998) and has the radial velocity companion with the largest value of *M sin i*, 11 $M_{Jup}$ (Latham et al. 1989). The true companion mass depends upon the orbital inclination which is difficult to assess. It is possible to measure the inclination of the stellar rotational axis from the rotational velocity of HD 114762; this also provides an estimate of the orbital inclination if the two axes are assumed to be aligned. Two studies have measured the projected stellar rotational velocity *v sin i* to be very low, ranging from 0-<1 km/s (Cochran et al. 1991) to 0.8±0.7 km/s (Hale 1995). With slightly different arguments about the true stellar velocity, both studies infer a low inclination angle for the stellar rotation axis (<11.5–20deg), assume approximate coplanarity between the stellar spin and the orbital axis, and suggest that the companion mass is actually near or above the hydrogen-burning limit. An analysis of Hipparcos astrometric data has also suggested that the companion may be an M star, but the short orbital period of the HD 114762 system complicates the interpretation of the Hipparcos measurements (Halbwachs et. al 2000)

The presence of a companion star only 130AU distant, however, may significantly affect the alignment between the stellar rotational axis and the orbital plane axis. An investigation of binary and triple systems with known orbits has measured the angle between the primary spin axis and the orbital plane axis of either the binary or the close pair in triples (Hale 1994). Based on these results, the primary spin is typically aligned (within 20 deg) with the orbital axis for close binaries, but is preferentially *misaligned* (average difference about 40 deg) in triple systems (Hale 1994), suggesting the presence of a wider companion may disrupt this correlation. For the HD 114762 system, an orbital axis inclined 40 deg from the largest measured stellar axis of 20 deg. could result in a value for *sin i* (orbital) as high as 0.87; this extreme case would imply a companion mass of $13M_{Jup}$, the boundary between planet and brown dwarf masses.

5.2 Sample

The binaries resolved by this project provide information about how a companion star influences planet formation. The small sample size for this investigation precludes a statistically significant analysis of the proportion of binaries among the stars with extrasolar planets, but a preliminary comparison is possible. The frequency of companions is calculated by the companion star fraction (*CSF*) defined as:

$$CSF=(b+2t)/(s+b+t)$$

where *s* is the number of singles, *b* is the number of binaries, and *t* is the number of triples.

Considering only the 0".15 to 10".0 separation range entirely covered by this dataset, the Tau Boo and HD 114762 systems result in a companion star fraction *CSF* $_{planet\ stars}$ = *0.18 ± 0.13*. The corresponding value for solar neighborhood G-dwarfs (Duquennoy & Mayor 1991) is *CSF* $_{field}$ = *0.27 ± 0.04*, assuming a distance of 15pc, a semi-major axis to observed separation of 1.26 (Fischer & Marcy 1992), and a total binary system mass of 1.4 $M_{\odot}$ (logP[dy] from 2.9 to 5.9). The proportion of binaries among stars with planets is similar to that of the field, especially considering the sample selection effect of no previously known <2" systems.

The results from a Keck AO companion search to stars with planets (Luhman & Jayawardhana 2002) can be incorporated to increase the sample size. For this study, the detection limits are similar and reach the bottom of the Main sequence at 3-10 AU, compared to 5-20 AU for our study. The Keck AO survey covered 25 extrasolar planetary systems, of which 8 are in common with our sample. The angular separation range coverage only extends to 3".3 for the Keck images, but the 17 stars not included in our data have a larger average distance of 34 pc which corresponds to an average 113 AU maximum projected separation; this is close to the value of 150 AU for our sample, though the 134 AU companion to HD 114762 would probably have been missed had that target been included in the Keck AO sample. The one Keck AO target with a significantly different separation range coverage (5 AU – 10 AU) is ε Eridani since it is such a nearby star. For this star, direct images from Keck using NIRC with an occulting finger (McCarthy 2001) are sensitive to companions from 15 AU – 115 AU; no possible stellar companions were detected and two potential substellar companions were ruled out by proper motion. Adding only the 3 stars that were sensitive to the entire 2 – 150 AU range of our sample lowers the proportion of companions to *CSF $_{planet\ stars}$ = 0.14 ± 0.10,* while including the null result for the entire 17 new targets further lowers the multiplicity to *CSF $_{planet\ stars}$ = 0.07 ± 0.05*. For the later case, the difference from the field G-dwarfs is a 2 σ result. Expanded searches including more of the growing sample of extrasolar planetary systems and covering a wide range of separations will better quantify the comparison of the *CSF* for stars with and without planets.

Binaries outside the field-of-view of the current images are included in catalogues, and these systems are listed in Appendix B. Among the 11 stars in the current sample, 5 have ADS (Aitken 1932) or IDS (Jeffers et. al 1963) double star designations, but only Tau Boo and 16 Cyg appear to be physically associated (see Appendix B). Additionally, a wide companion (55" or ~750 AU) to Ups And has recently been discovered with multi-epoch images (2MASS, POSS-I and POSS-II) and follow-up spectroscopy (Lowrance et al. 2002). Completeness is less certain for this larger range (extending to the 55" separation of Ups And), but counting the previously-known binaries Tau Boo and 16 Cyg A/B along with the newly-discovered companions to HD 114762, 16 Cyg A, and Ups And gives a value of *CSF $_{planet\ stars}$ = 0.45 ± 0.20* over a wider separation range from 0".15 to 55". With the same assumptions about distance and total binary system mass, the G-dwarf *CSF* for logP[dy] from 2.9 to 7.2 is *CSF* $_{field}$ = *0.41 ± 0.05.*

Although the statistics are poor, the binary fraction resembles that of the field, suggesting that binaries may not be as hostile to planet formation at separations <few AU, despite diminished disk signatures in young binaries (e.g. Jensen et al. 1996). The results are more consistent with the results of an *HST* study of young solar-type binary stars with projected separations <50 AU which showed circumstellar accretion disk signatures at levels indistinguishable from young single stars (White & Ghez 2001).

Observational trends associating the binary parameters and the planetary properties also relate to planet formation and evolution models. One simulation of planet formation predicts that planets will not form in equal mass binaries with separations <50 AU (Nelson 2000). The Tau Boo system has a companion at a projected separation of 45 AU, within the range investigated by the simulation. Although the presence of this companion did not prevent a planet from forming very close to Tau Boo, this system does not necessarily contradict the simulation results since the binary has unequal mass components – the companion is an M star, while the primary is a late-F star. Observations of a larger sample of stars with planets will better characterize the mass ratio limit of binaries with planets. Since the current radial velocity searches avoid stars with known companions within 2", encompassing much of the range of the simulation, it is difficult to test the hypothesis. The combination of adaptive optics and spectroscopy demonstrated in this work, however, will lead to surveys capable of monitoring the components of binaries separated by <2".

Another prediction concerning binaries is the expectation that a companion star will perturb the planetary orbit, making it eccentric (e.g. Mazeh et al. 1997, Artymowicz 1998). Among the stars observed for this project, the HD 114762 and 16 Cyg B have planets in eccentric orbits. The third binary Tau Boo has a planet with a tidally-circularized orbit for which it is impossible to know the initial orbital eccentricity. These three systems do not account for all the eccentric planetary orbits, however, and there may be additional substellar companions to the remaining high-eccentricity planet stars which would be undetectable with the current data. The only non-circularized, low-eccentricity planet found in a catalogued binary system is HD 195019 (Aitken 1932). Since this system is not included in the more recent WDS binary catalogue, future observations are planned to confirm that this is an associated binary. Unfortunately, the current sample is still too small to determine if all planets orbiting binary stars are driven to eccentric orbits.

6. Summary

Among the 11 stars with radial velocity planets observed in this study, three companions were detected – a previously unknown companion to HD 114762, a new component to the known pair 16 Cyg A/B, and the catalogued secondary to Tau Boo. Considering the 0".15 to 10".0 separation range covered for the entire sample, the companion star fraction for the stars with planets is $CSF_{planet\ stars} = 0.18 \pm 0.13$ which is not significantly different from the value for field G-dwarfs $CSF_{field} = 0.27 \pm 0.04$. Although based on a small sample, this similarity in binary frequency suggests that companions – even when separated by <50 AU – do not entirely preclude planet formation, at least not the close, massive planets detected by radial velocity surveys. The presence of planets in some binary systems also suggests that it may be worthwhile to include known binaries in planet search programs. Two of the three multiple systems observed contain planets in eccentric orbits, and the third binary has a planet in a tidally-circularized orbit. Several single stars in the observed sample also have eccentric planets, however.

The observations of the binaries Tau Boo and HD 114762 relate to specific questions about these two systems. A recent infrared spectroscopic investigation (Wiedemann et. al 2000) has led to the possibility that there is an additional substellar companion to Tau Boo, but the interpretation of the data depends upon the contamination by the secondary through the slit. Based on the current data, it is unlikely that light from the secondary contributed significantly to the spectrum. Additionally, the orbital parameters of the Tau Boo system are uncertain (Hale 1994, Popovic & Pavlovic 1996) and the current data will help empirically measure the periastron time and separation. The periastron separation and orbit eccentricity are important to determine since numerical simulations of disks perturbed by a binary companion or a stellar fly-by (Artymowicz & Lubow 1994, Clarke & Pringle 1993) suggest that some combinations of these parameters would have truncated circumstellar disks to radii near the ice condenstation radius. Finally, the companion to HD 114762 is an important factor to consider in the estimation of the inclination angle of the companion; for this system the inclination angle is especially important in establishing a substellar mass for the companion. Given the new information about the companion star, the low inclination angle for the primary actually makes a high inclination angle for the substellar companion possible, based on statistics of inclination angles compiled for triples (Hale 1994).

Acknowledgements

We gratefully acknowledge the substantial help from the staff at Lick and Keck observatories, particularly support scientist Elinor Gates, Lick telescope operators Keith Baker, Jim Burrous, Kostas Chloros, Wayne Earthman, John Morey, and Andy Tulles, and Keck telescope operators Joel Aycock and Ron Quick. This work was performed under the auspices of the U. S. Department of Energy, National Nuclear Security Administration by the University of California, Lawrence Livermore National Laboratory under Contract W-7405-ENG-48. Some of the data for this project are based on observations at the W. M. Keck Observatory, which is operated as a scientific partnership between Caltech, the University of California, NASA, and made possible by the generous financial support of the W. M. Keck Foundation. The authors wish to recognize and acknowledge the very significant cultural role and reverence that the summit of Mauna Kea has always had within the indigenous Hawaiian community. We are most fortunate to have the opportunity to conduct observations from this mountain. Support for this work was provided by NASA through grant number NAG5-6975 under the Origins of Solar Systems Program. This work has also been supported in part by the National Science Foundation Science and Technology Center for Adaptive Optics, managed by the University of California at Santa Cruz under cooperative agreement No. AST-9876783. This research required information from the SIMBAD and Vizier databases, operated at CDS, Strasbourg, France.

References

Aitken, R. G. 1932, New General Catalogue of Double Stars (Washington: Carnegie Inst. of Washington).

Artymowicz, P. 1998 in *Brown dwarfs and extrasolar planets, Proceedings of a Workshop held in Puerto de la Cruz, Tenerife, Spain 17-21 March 1997*; ASP Conference Series #134 eds. Rafael Rebolo; Eduardo Martin; Maria Rosa Zapatero Osorio.

Artymowicz, P. & Lubow, S. H. 1994, ApJ, 421, 651.

Baliunas, S. L., Henry, G. W., Donahue, R. A., Fekel, F. C., & Soon, W. H. 1997, ApJ, 474, L119.

Bauman, B. J., Gavel, D. T., Waltjen, K. E., Freeze, G. J., Keahi, K. A., Kuklo, T. C., Lopes, S. K., Newman, M. J., Olivier, S. S. 1999 Proc. SPIE, 3762, 194.

Bodenheimer, P. & Pollack, J. B. 1986, Icarus, 67, 391.

Boss, A. P. 1995, Science, 267, 360.

Butler, R. P. & Marcy, G. W. 1996, ApJ, 464, L153

Butler, P., Marcy, G., Fischer, D., Brown, T., Contos, A., Korzennik, S., Nisenson, P., & Noyes, R. 1999, ApJ, 526, 916.

Butler, R. P., Marcy, G. W., Williams, E., Hauser, H., & Shirts, P. 1997, ApJ, 474, L115.

Cameron, A. G. W. 1978, M&P, 18, 5.

Clarke, C. J. & Pringle, J. E. 1993, MNRAS, 261, 190.

Cochran, W. D., Hatzes, A. P., Hancock, T. J. 1991, 380, L35.

Cochran, W. D., Hatzes, A. P., Butler, R. P., & Marcy, G. W. 1997, ApJ, 483, 457.

Duquennoy, A. & Mayor, M. 1991, A&A, 248, 485

Fischer, D. A. & Marcy, G. W. 1992, ApJ, 396, 178.

Fischer, D. A., Marcy, G. W., Butler, R. P., Vogt, S. S., Frink, S., & Apps, K. 2001, ApJ, 551, 1107.

Fischer, D. A., Marcy, G. W., Butler, R. P., Laughlin, G., & Vogt, S. 2002, ApJ, 564, 1028.

Gavel, D. T., Bauman, B. J., Campbell, E. W., Carrano, C. J., & Olivier, S. S. 1999 Proc. SPIE, 3762, 266.

Ghez, A. M., Neugebauer, G. & Matthews, K. 1993, AJ, 106, 2005

Ghez, A. M., McCarthy, D. W., Patience, J., & Beck, T. L. 1997, ApJ, 481, 378.

Ghez, A. M., Klein, B. L., Morris, M., & Becklin, E. E. 1998, ApJ, 509, 678.

Gilmore, D. K., Rank, D. M., & Temi, P. 1994, in Instrumentation in Astronomy VIII, ed. D. L. Crawford, & E. R. Crane (Proc. SPIE , Vol. 2198), 744.

Halbwachs, J. L., Arenou, F., Mayor, M., Udry, S. & Queloz, D. 2000, A&A, 355, 581.

Hale, A. 1995, PASP, 107, 22.

Hale, A. 1994, AJ, 107, 306.

Heintz, W. D. 1971, *Double Stars*, D. Reidel Publishing Company

Henry, G. W., Baliunas, S. L., Donahue, R. A., Soon, W. H., & Saar, S. H. 1997, ApJ, 474, 503.

Henry, T. J. & McCarthy, D. W. 1993, AJ, 106, 773.

Jeffers, H. M., van den Bos, W. H., & Greeby, F. M. 1963, *Index Catalogue of Visual Double Stars* 1961.0 (Publ. Lick Obs., No. 21) (Mt. Hamilton: Univ. of California).

Jensen, E. L., Mathieu, R. D., & Fuller, G. A. 1996, ApJ, 458, 312.

Labeyrie, A. 1970 A&A, 6, 85.

Latham, D. W., Mazeh, T., Stefanik, R. P., Mayor, M., & Burki, G. 1989, Nature, 339, 38L.

Lloyd, J. P., Liu, M. C., Macintosh, B. A., Severson, S. A., Deich, W. T. S., & Graham, J. R. 2000, Proc. SPIE, 4008, 814.

Lohmann, A. W., Weigelt, G., & Wirtzner, B. 1983, Appl. Opt., 22, 4028.

Lowrance, P. J., Kirkpatrick, J. D., & Beichman, C. 2002, ApJL, 572, 79.

Luhman, K. L. & Jayawardhana, R. 2002, ApJ, 566, 1132.

Marcy, G. W. & Butler, R. P. 1996, ApJ, 464, L147

Marcy, G. W., Butler, R. P., Williams, E., Bildsten, L., Graham, J. R., Ghez, A. M., & Jernigan, J. G. 1997, ApJ, 481, 926.

Marcy, G. W., Butler, R. P., Vogt, S. S., Fischer, D., Lissauer, J. J. 1998, ApJ, 505, L147.

Marcy, G. W., Cochran, W. D., & Mayor, M. 1998 in Protostars & Planets IV ed. V. Mannings (Tuscon: Univ. Arizona Press).

Marcy, G. W., Butler, R. P., Fischer, D., Vogt, S. S., Lissauer, J. J., & Rivera, E. J. 2001, ApJ, 556, 296.

Marcy, G. W., Butler, R. P., Fischer, D. A., Laughlin, G., Vogt, S. S., Henry, G. W., & Pourbaix, D. 2002, ApJ, submitted.

Matthews, K.& Soifer, B. T. 1994, Astronomy with Infrared Arrays: The Next Generation, ed. I. McLean (v. 190; Dordrecht: Kluwer), 239.

Matthews, K., Ghez, A. M., Weinberger, A. J., & Neugebauer, G. 1996, PASP, 108, 615.

Mayor, M. & Queloz, D. 1995 Nature, 378, 355.

Mazeh, T., Krymolowski, Y., Rosenfeld, G. 1997, ApJ, 477, L103.

McCarthy, C. 2001, PhD thesis, UCLA.

McLean et al. 1998, Proc. SPIE, 3354, 566.

McLean et al. 2000, ApJ, 533, L45.

Nelson, A. F. 2000 ApJ, 537, L65.

Noyes, R. W., Jha, S., Korzennik, S. G., Krockenberger, M., Nisenson, P., Brown, T. M., Kennelly, E. J., & Horner, S. D. 1997, ApJ, 483L, 111.

Popovic, G.M & Pavlovic, R. 1996, Bull. Astron. Belgrade 153, 57.

Reid, I. N., Burgasser, A. J., Cruz, K. L., Kirkpatrick, J. D., & Gizis, J. E., 2001, AJ, 121, 1710.

Rudy. R. J., Woodward, C. E., Hodge, T., Fairfield, S. W., & Harker, D. E. 1997, Nature, 387, 159.

Schneider, G., Becklin, E. E., Smith, B. A., Weinberger, A. J., Silverstone, M., & Hines, D. C. 2001, AJ, 121, 1163.

Simon, M., Ghez, A. M., Leinert, Ch., Cassar, L., Chen, W. P., Howell, R. R., Jameson, R. F., Matthews, K., Neugebauer, G. & Richichi, A. 1995, ApJ, 443, 625

Shu, F. H., Adams, F. C., & Lizano, S. 1987 ARAA, 25, 23.

Trilling, D. E. & Brown, R. H. 1998, Nature, 395, 775.

White, R. J. & Ghez, A. M. 2001, ApJ, 556, 265.

Wiedemann, G., Deming, D., & Bjorker, G. 2001, ApJ, 546, 1068.

Wizinowich, P., Acton, D. S., Shelton, C., Stomski, P., Gathright, J., Ho, K., Lupton, W., Tsubota, K., Lai, O., Max, C., Brase, J., An, J., Avicola, K., Olivier, S., Gavel, D., Macintosh, B., Ghez, A., & Larkin, J. 2000, PASP, 112, 315.

Worley, C. E. & Douglass, G. G. 1997, A&AS, 125, 523.

Appendix A

Combining the 2.2 μm data from each star, the median detection limits for each technique with more than one observation – Keck speckle, Keck shift-and-add, Lick AO imaging, and Keck direct imaging – are shown in Figure A1. Two curves are plotted for Lick AO data since two cameras were used and the current camera IRCAL is significantly more sensitive than the initial camera LIRCII. For most targets, Keck speckle data defines the detection limit at subarcsecond separations, while Lick AO data sets the detection threshold at separations exceeding an arcsecond. The individual limits from each target are given in Figures A2 – A13. The individual limits are again differentiated by technique.

Appendix B

Although the entire set of stars with extrasolar planets has not been systematically observed for this project, all the systems with companions listed in the ADS (Aitken 1932) or IDS (Jeffers et al. 1963) catalogues are included in Table A1. As planets are found around fainter primary stars, it is less likely that these earlier catalogues are complete, but the currently known multiple systems are listed here for reference. Some of the systems are only listed in older catalogues and not the more recent WDS (Worley & Douglass 1997), making it unlikely that these are associated pairs. Many of these systems are very wide doubles and would not have been detected by this survey; Tau Boo and Upsilon And are the only observed binaries with separations within the field-of-view of the wide field cameras. The proper motion of the companion to Upsilon And is not given in SIMBAD, but, because it was not seen in our wide-field data, it is probably a background object; only one measurement is given in the IDS catalogue. Another double, rho CrB, does have multiple measurements which show the second star moving a substantial amount from the target despite a large separation; this is not a common proper motion pair. 70 Vir is not listed in the WDS catalogue and is probably a chance superposition also. Among the 11 stars in this sample, only Tau Boo and 16 Cyg B are confirmed binaries with either orbital solutions or measured common proper motion. The additional two systems not included in the current observations include one physically associated binary system – HD 195019.

Tables

Table 1: Sample

| Target | RA (2000) | Dec (2000) | p.m. RA ("/yr) | p.m. Dec ("/yr) | V | SpTy | Ref. | D (pc) | Msini (Mjup) | a (AU) | Ecc. | Residual | Ref. |
|---|---|---|---|---|---|---|---|---|---|---|---|---|---|
| Ups And | 01 36 47.8 | +41 24 19.7 | -0.1726 | -0.3810 | 4.09 | F8V | a | 13.5 | 0.63 | 0.053 | 0.1 | Yes | 1, 11 |
| 55 Cnc | 08 52 35.8 | +28 19 50.9 | -0.485 | -0.2344 | 5.95 | G8-K0V | a | 12.5 | 0.85 | 0.12 | 0.04 | Yes | 1, 11 |
| 47 UMa | 10 59 28.0 | +40 25 48.9 | -0.3159 | -0.0552 | 5.10 | G1V | b | 14.1 | 2.42 | 2.08 | 0.03 | Yes | 2, 11 |
| HD 114762 | 13 12 19.7 | +17 31 01.6 | -0.583 | -0.0020 | 7.30 | F9V | b | 40.6 | 10 | 0.4 | 0.25 | | 3 |
| 70 Vir | 13 28 25.8 | +13 46 43.6 | -0.2348 | -0.5762 | 5.00 | G4V | b | 18.1 | 6.84 | 0.47 | 0.40 | | 4, 11 |
| Tau Boo | 13 47 15.7 | +17 27 24.9 | -0.4803 | 0.0542 | 4.50 | F7V | a | 15.6 | 3.64 | 0.042 | 0.00 | Yes | 1, 11 |
| Rho CrB | 16 01 02.7 | +33 18 12.6 | -0.1969 | -0.7730 | 5.40 | G0Va | c | 17.4 | 1.1 | 0.23 | 0.03 | | 5, 11 |
| 14 Her | 16 10 24.3 | +43 49 03.5 | 0.1325 | -0.2984 | 6.67 | K0V | c | 18.1 | 4 | 3 | 0.35 | | 6 |
| 16 Cyg B | 19 41 52.0 | +50 31 03.1 | -0.1478 | -0.1589 | 6.20 | G3V | c | 21.4 | 1.74 | 1.7 | 0.69 | | 7, 11 |
| Gl 876 | 22 53 16.7 | -14 15 49.3 | 0.960 | -0.676 | 10.17 | M4 | c | 4.70 | 2.1 | 0.21 | 0.27 | Yes | 8, 11 |
| 51 Peg | 22 57 28.0 | +20 46 07.8 | 0.2081 | 0.0610 | 5.49 | G5V | b | 15.4 | 0.44 | 0.051 | 0.01 | | 9,10, 11 |

Coordinates are epoch 2000; Multiple planets are detected around Ups And (Butler et al. 1999), 55 Cnc (Marcy et al. 2002), 47 Uma (Fischer at al. 2002), and Gl 876 (Marcy et al. 2001); a – Baliunas et al. 1997; b – Henry et al. 1997; c – SIMBAD database; 1 – Butler et al. 1997; 2 – Butler & Marcy 1996; 3 – Latham et al. 1989; 4 - Marcy & Butler 1996; 5 – Noyes et al. 1997; 6 – Marcy et al. 1998; 7 – Cochran et al. 1997; 8 – Marcy et al. 1998; 9 – Mayor & Queloz 1995; 10 – Marcy et al. 1997; 11 – Fischer et al. 2001

Table 2: Imaging Instruments

| Camera | Telescope | Pixel Scale | Field-of-view | Technique | Ref |
|---|---|---|---|---|---|
| NIRC + Scale changer | Keck I | 0".021 | 5".3 x 5".3 | Speckle | 1, 2 |
| NIC | Lick 3m | 0".065 | 4".2 x 4".2 (subarray) | Speckle | 3 |
| IRCAL | Lick 3m + AO | 0".075 | 19" x 19" | AO Imaging | 4 |
| LIRC II | Lick 3m + AO | 0".125 | 32" x 32" | AO Imaging | 5 |
| NIRC | Keck I | 0".150 | 38" x 38" | Direct Imaging | 1 |

1 – Matthews & Soifer 1994; 2 – Matthews et al. 1996; 3 – Rudy et al. 1997; 4 Lloyd et al. 2000; 5 – Gilmore et al. 1994

Table 3: Summary of Observations

| Target | Camera | Filter | Date | Type |
|---|---|---|---|---|
| Upsilon And | LIRC II | *Brγ, K'* | 98 Sep 7 | High-res, Wide field |
| | NIRC + scale changer | *Brγ* | 98 Nov 26 | High-res |
| | NIRC | *Brγ* | 98 Nov 26 | Wide field |
| 55 Cnc | NIRC + scale changer | *Brγ* | 97 Jan 2 | High-res, |
| 47 UMa | NIRC + scale changer | *Brγ* | 97 Jan 2 | High-res. |
| HD 114762 | NIRC + scale changer | *Brγ* | 97 Jan 2 | High-res. |
| | NIRC | *Brγ, K, OII* | 98 Nov 25 & 98 Nov 26 | Wide field |
| | IRCAL | *Brγ, J&H w/finger* | 00 May 21 | High-res, Wide field |
| 70 Vir | NIRCAM | $K_s$ | 97 June 25 | High-res. |
| | IRCAL | *Brγ, $K_s$* | 20 June 00 | High-res, Wide field |
| Tau Boo | NIRC + scale changer | *Brγ* | 98 Nov 26 | High-res |
| | NIRC | *Brγ, OII* | 98 Nov 26 | Wide field |
| | IRCAL | *J, H, $K_s$ (all w/ND)* | 16 June 00 | High-res, Wide field |
| Rho CrB | IRCAL | *Ks* | 00 Aug 2 | High-res, Wide field |
| 14 Her | IRCAL | *Brγ, $K_s$* | 01 Aug 13 | High-res, Wide field |

| | | | | | |
|---|---|---|---|---|---|
| 16 Cyg B | LIRC II | *Brγ, K'* | 98 Sep 7 | | High-res, Wide field |
| | NIRC + scale changer | *Brγ* | 98 Nov 25 | | High-res |
| | NIRC | *Brγ, K, J* | 98 Nov 25 | | Wide field |
| 16 Cyg A | NIRC | *Brγ* | 98 Nov 25 | | Wide field |
| | IRCAL | *H+ND* | 99 Aug 3 | | High-res, Wide field |
| | IRCAL | $K_s$ *+ND* | 00 May 21 | | High-res, Wide field |
| Gl 876 | NIRC + scale changer | *Brγ* | 98 Nov 26 | | High-res. |
| | NIRC | *Brγ* | 98 Nov 26 | | Wide field |
| 51 Peg | NIRC + scale changer | *Brγ* | | | High-res |
| | LIRC II | *Brγ, K'* | 98 Sep 7 | | High-res, Wide field |

Table 4: Measured Binary Parameters

| Binary | Date | Separation | Position Angle | Δmag | λ |
|---|---|---|---|---|---|
| HD 114762 | 98 Nov 26 | 3".26 ± 0".04 | 29.0 ± 0.1 | 7.35 ± 0.05 | *OII*=1.236 μm |
| | 98 Nov 25 | 3".33 ± 0".05 | 29.6 ± 0.9 | 7.3 ± 0.2 | *Brγ*=2.165 μm |
| HD 114762 | 00 June 16 | 3".26 ± 0".05 | 30.4 ± 1.0 | 7.6 ± 0.1 | *J + ND* |
| | | | | 7.5 ± 0.1 | *H + ND* |
| | | | | 7.2 ± 0.1 | $K_s$ *+ ND* |
| Tau Boo | 98 Nov 26 | 2".87 ± 0".03 | 30.8 ± 0.3 | 4.2 ± 0.1 | *OII* |
| | 98 Nov 26 | 2".86 ± 0".03 | 30.9 ± 0.5 | 3.1 ± 0.1 | *Brγ* |
| Tau Boo | 00 June 16 | 2".82 ± 0".04 | 33.2 ± 1.0 | 3.55 ± 0.05 | $K_s$*+ND* |
| 16 Cyg B | 98 Nov 25 | 15".77 ± 0".16 | 75.9 ± 0.1 | 7.49 ± 0.06 | *Brγ* |
| (unassociated) | 98 Nov 25 | Saturated | Saturated | 7.5 ± 0.1 | $K_s$ |
| | 98 Nov 25 | Saturated | Saturated | 7.9 ± 0.1 | *J* |
| 16 Cyg A | 98 Nov 25 | 3".44 ± 0".04 | 203.7 ± 0.2 | 5.10 ± 0.05 | *Brγ* |
| | 99 Aug 3 | 3".43 ± 0".05 | 205.0 ± 1.0 | 5.36 ± 0.1 | *H+ND* |
| | 00 May 21 | 3".43 ± 0".05 | 204.7 ± 1.0 | 5.34 ± 0.1 | *$K_s$ +ND* |

Table A1: Catalogued Binaries among Stars with Planets

| Target | Binary Name | Separation | Delta mag | Associated | Notes |
|---|---|---|---|---|---|
| Tau Boo | ADS 9025 | 4".5 | ΔV=6.6 | Yes | Orbital motion measured |
| Ups And | IDS01310+4054 | 14" | ΔV=8.4 | No | Not seen in NIRC or LIRCII image |
| | 2MASSI J0136504+412332 | 55" | ΔJ=6.30 | Yes | Common proper motion |
| HD 195019 | ADS 13886 | AB=3".7 | Δ=3.4 | Yes | Motion too small for background |
| | | AC=95".8 | Δ=2.9 | No | Not common proper motion |
| HD 192263 | ADS 13547 | AB=65".26 | Not given | No | Not common proper motion |
| | | AD=71".25 | Not given | No | Only 1 measurement |
| Rho CrB | IDS 15573+3337 | 89".6 | ΔV=3.2 | No | Not common proper motion |
| 70 Vir | IDS 13236+1419 | 86".3 | ΔV=3.6 | No | Not common proper motion |
| 16 Cyg B | ADS 12815 | 39".0 | ΔV=0.1 | Yes | Common proper motioin |

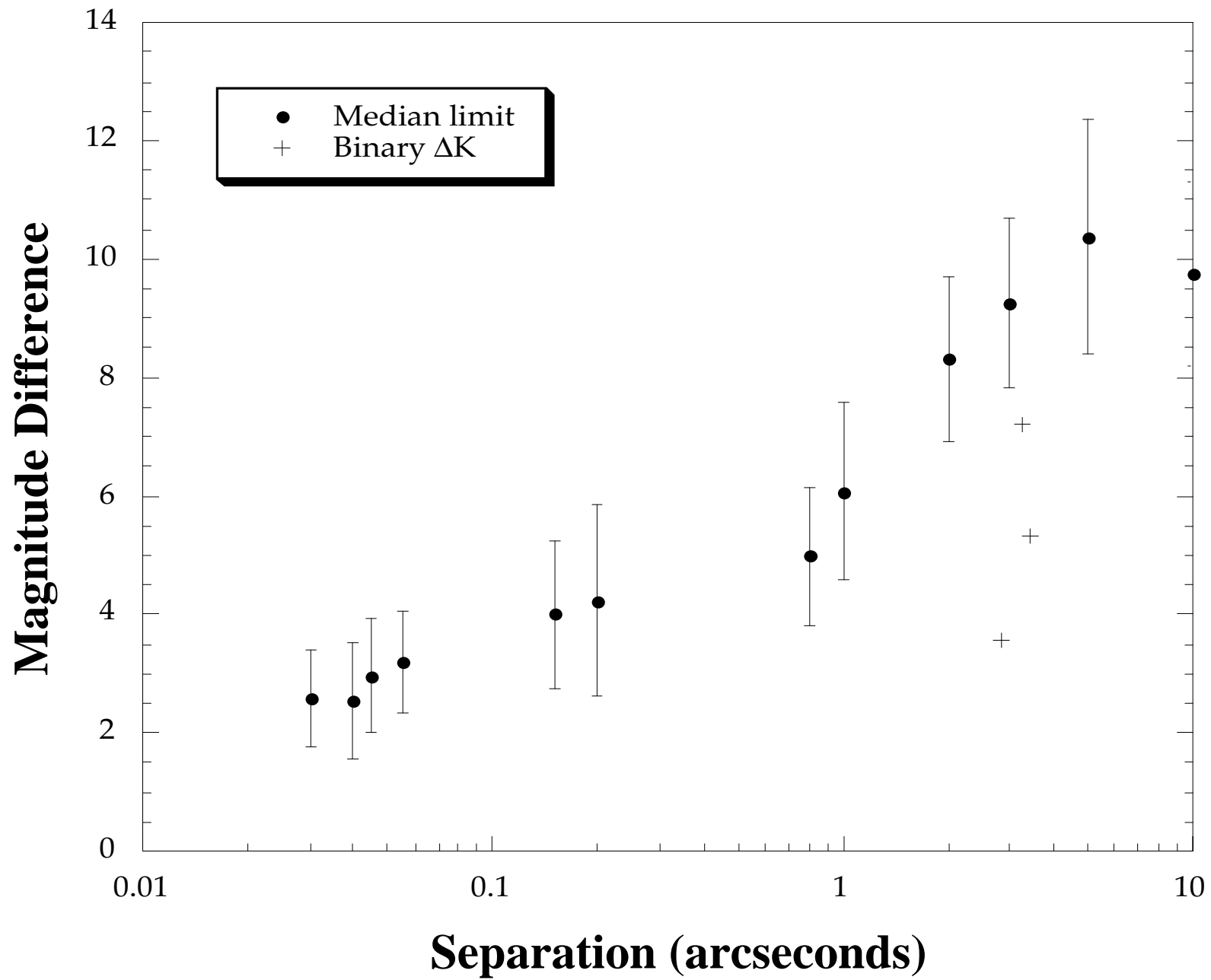

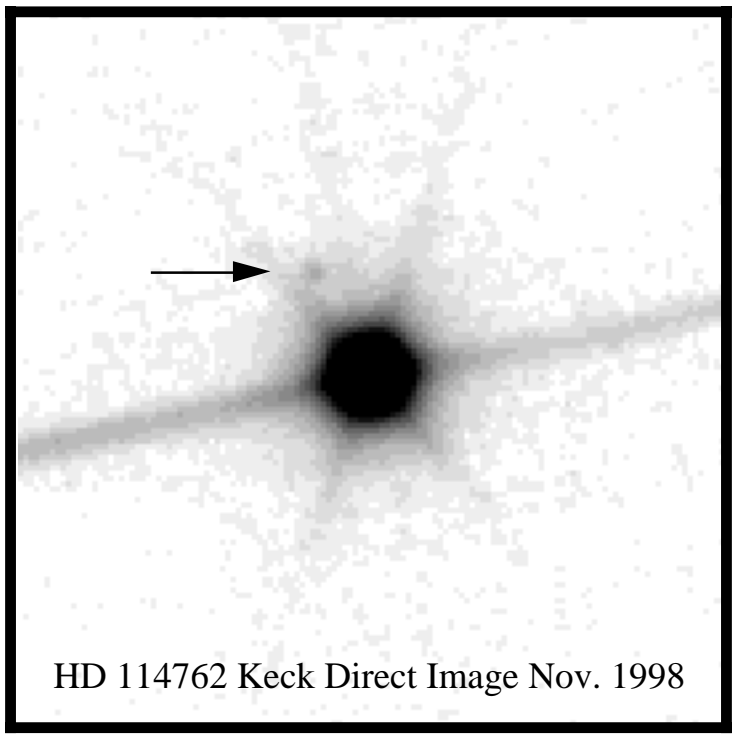 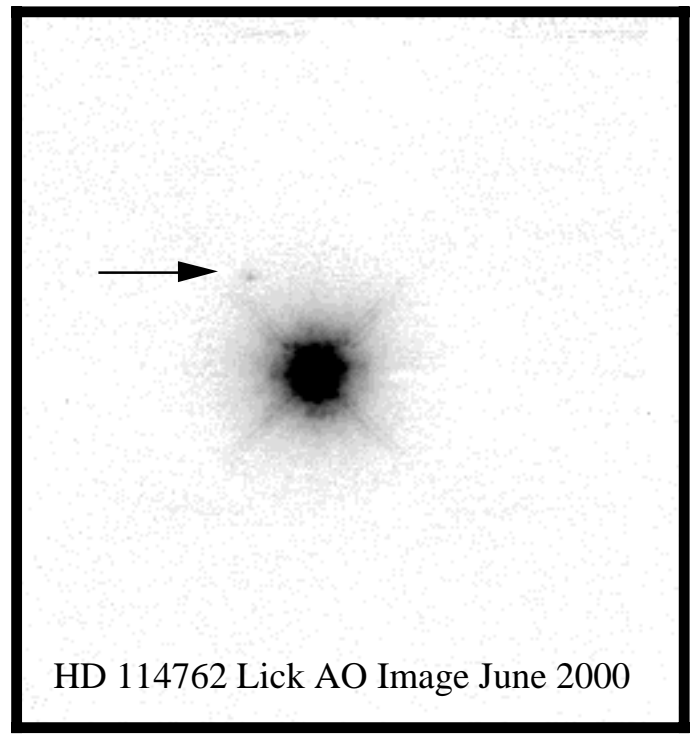

HD 114762 Keck Direct Image Nov. 1998    HD 114762 Lick AO Image June 2000

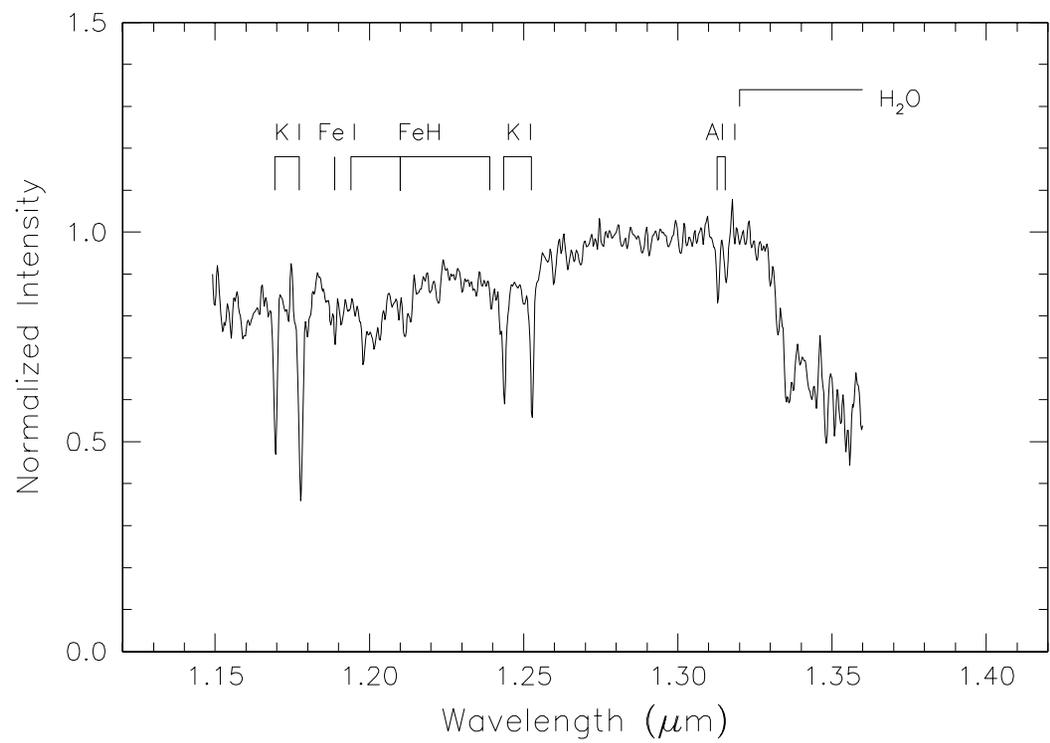

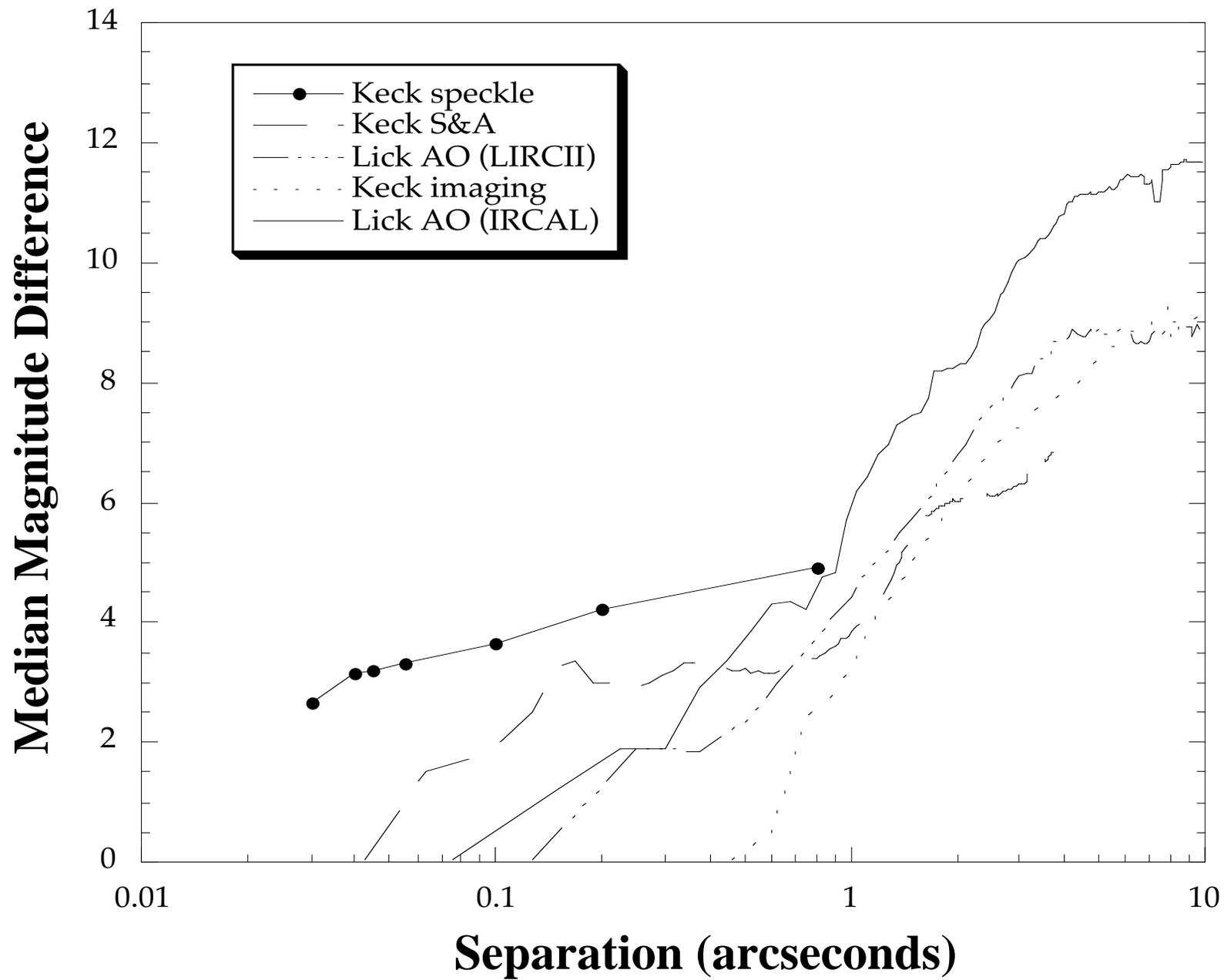

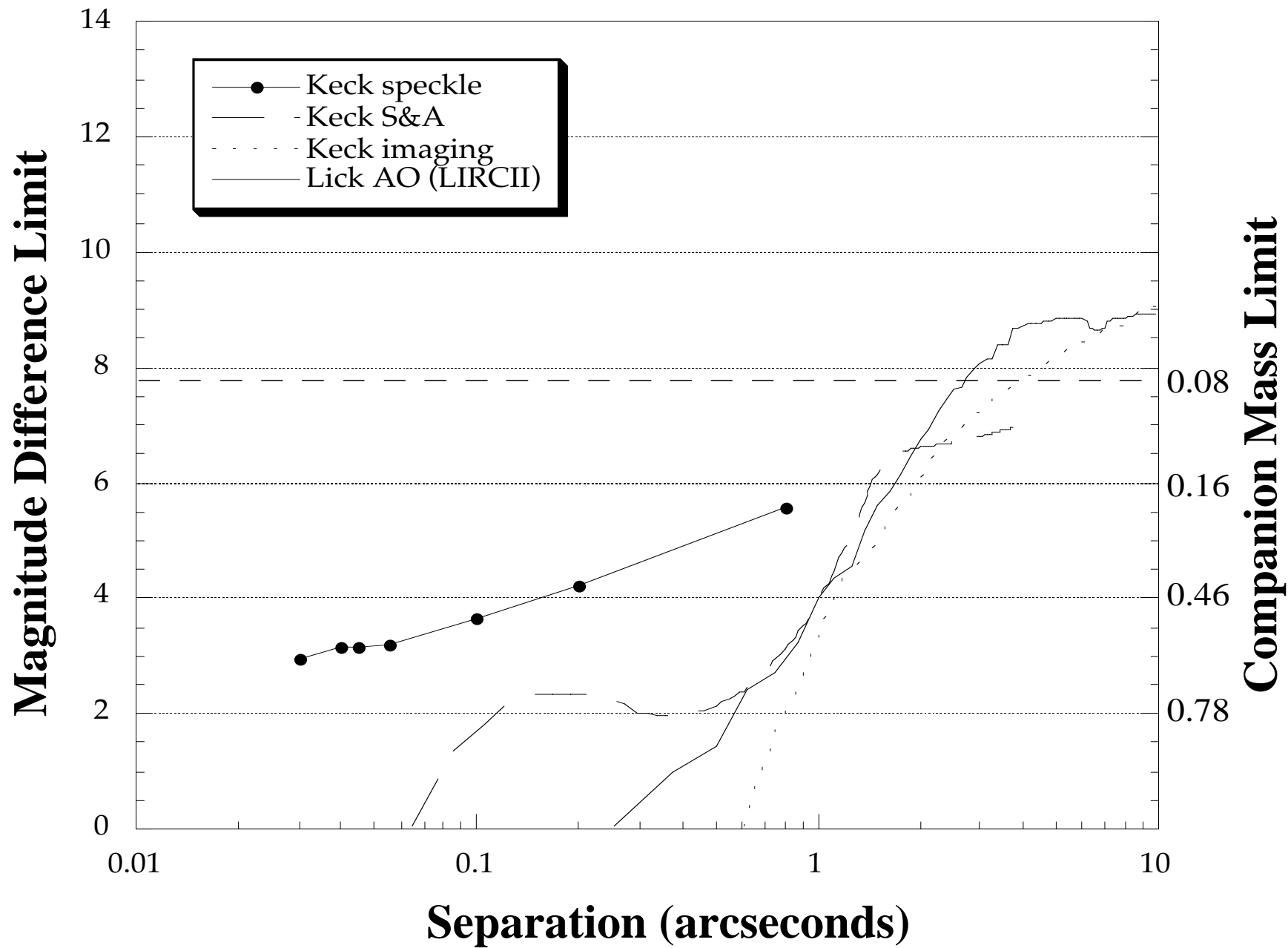

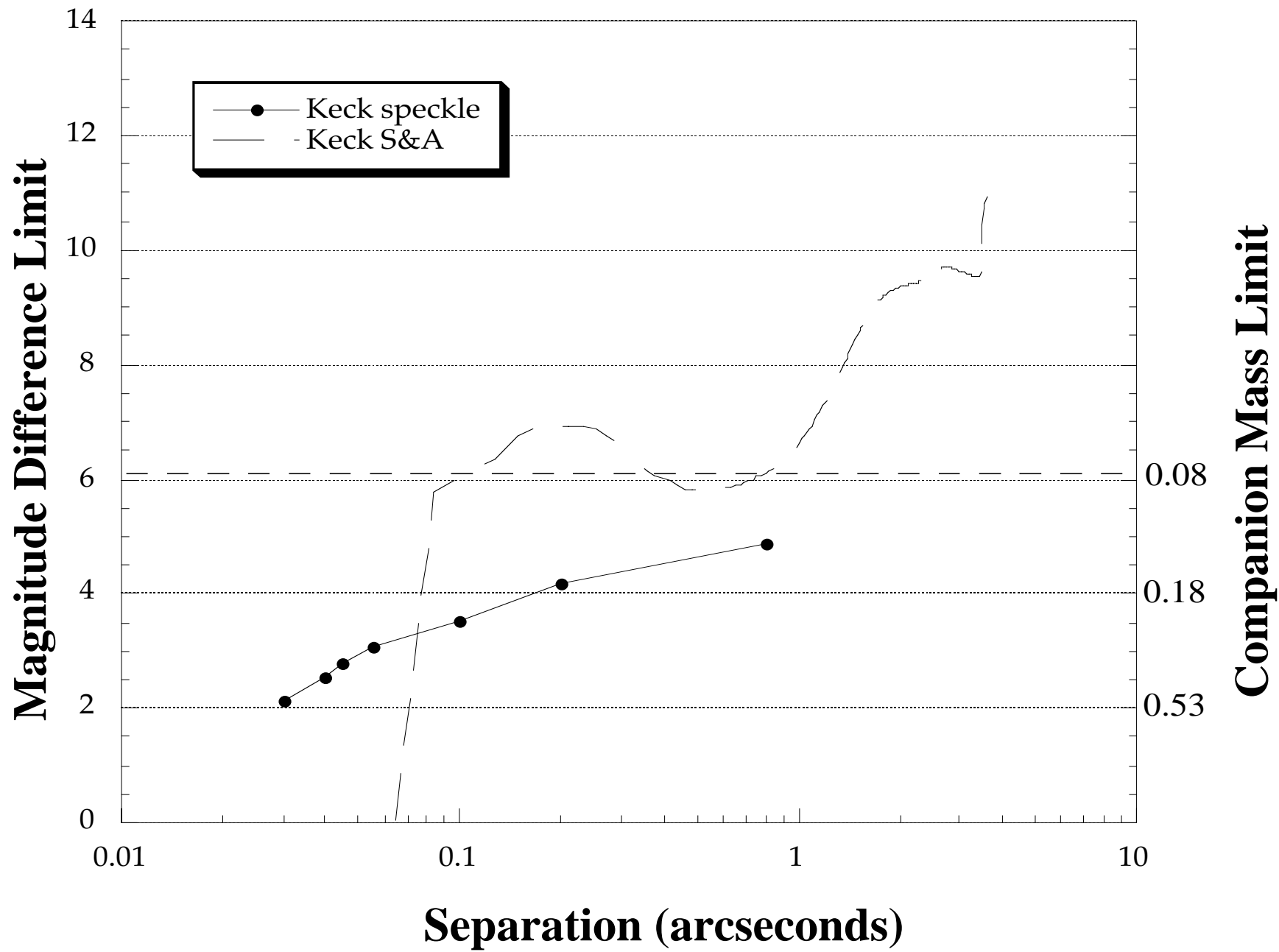

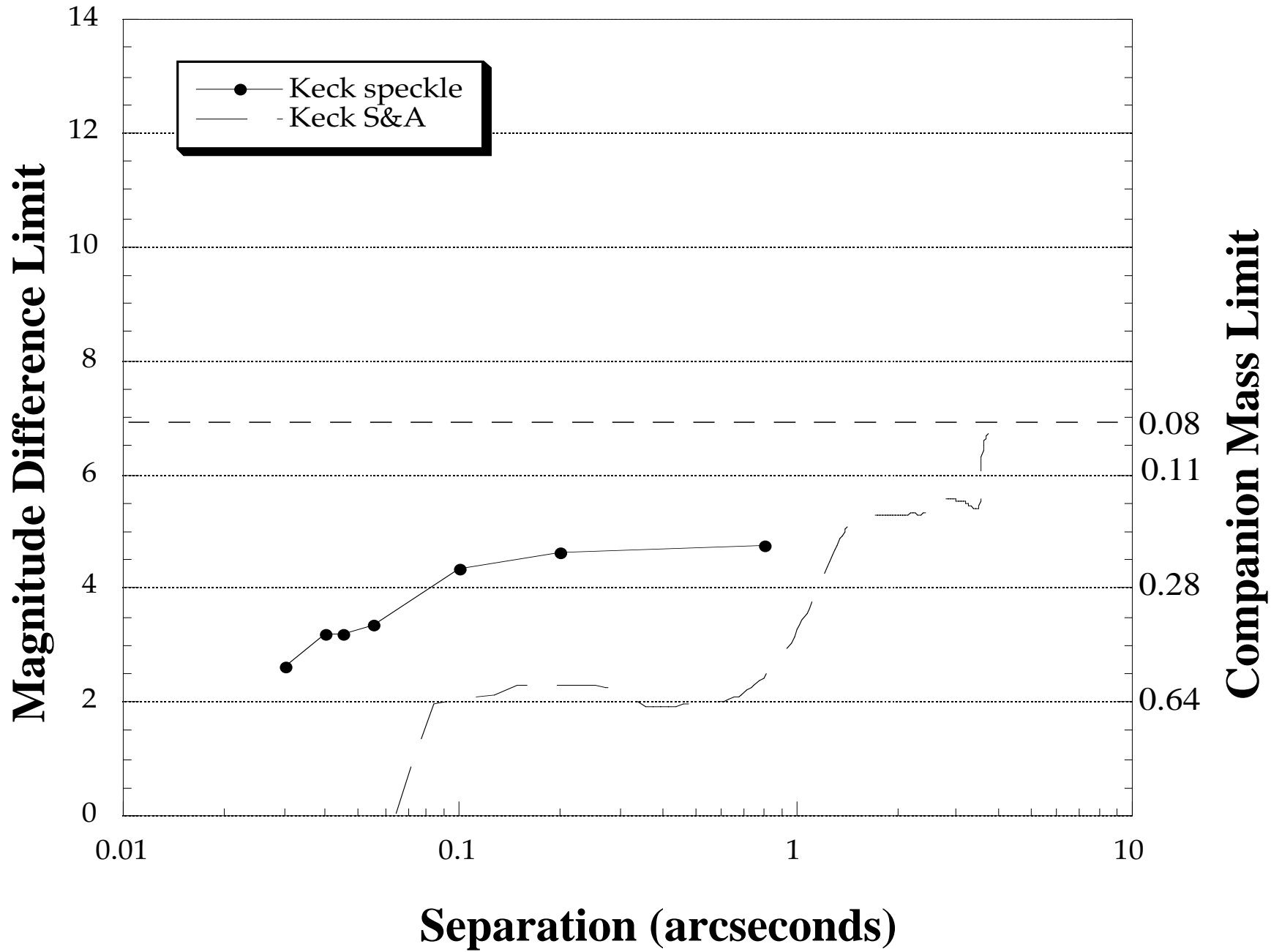

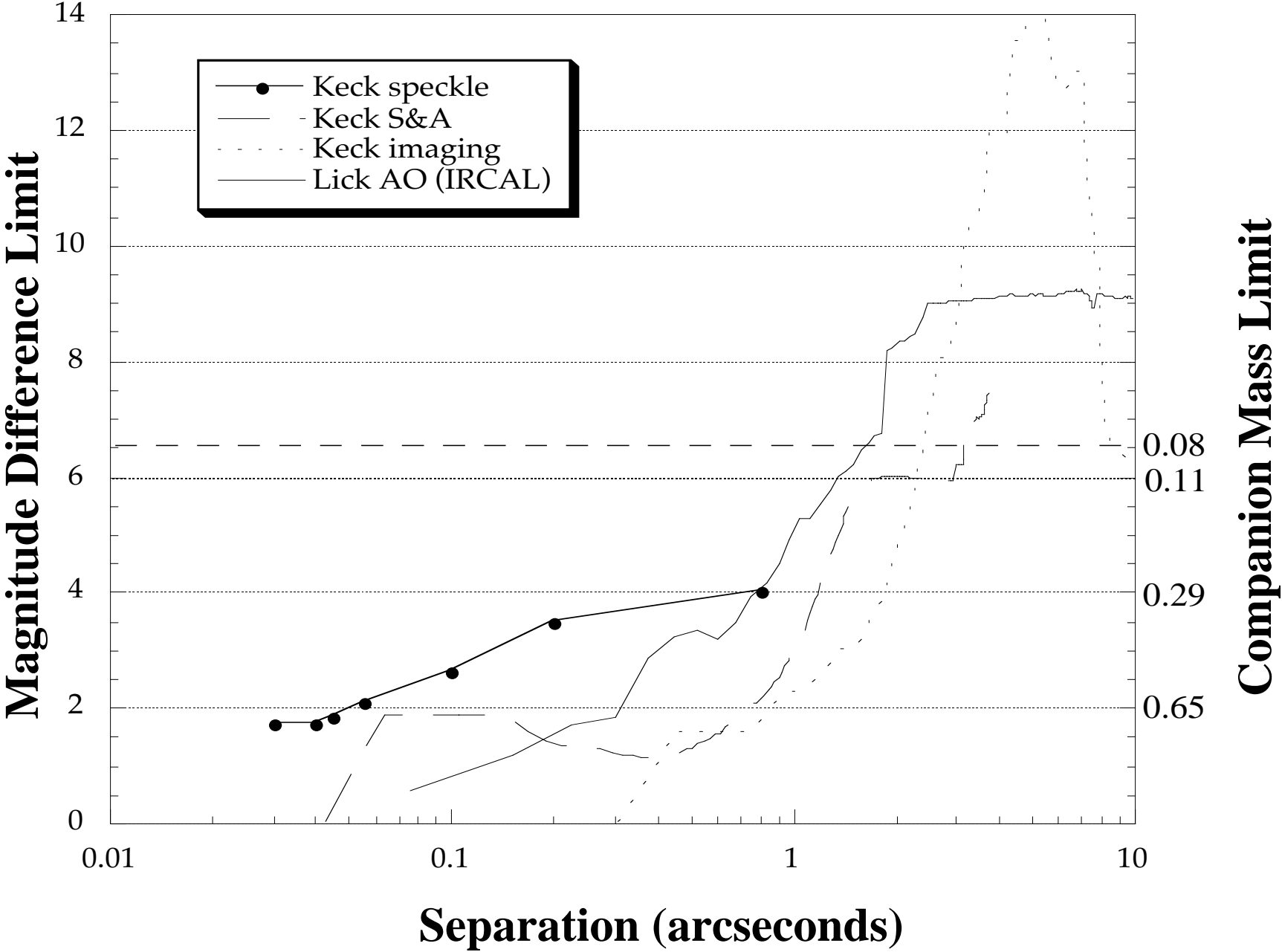

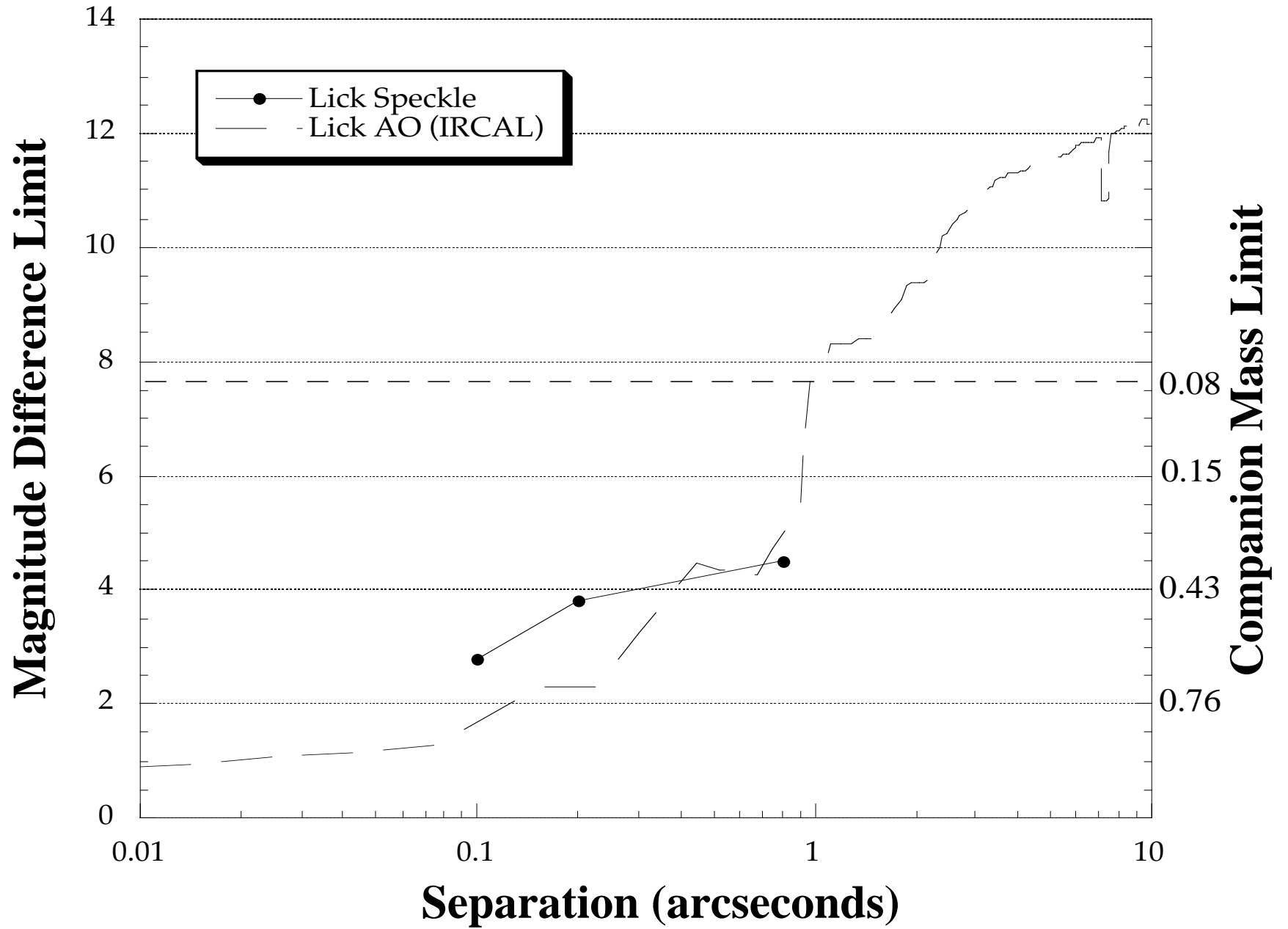

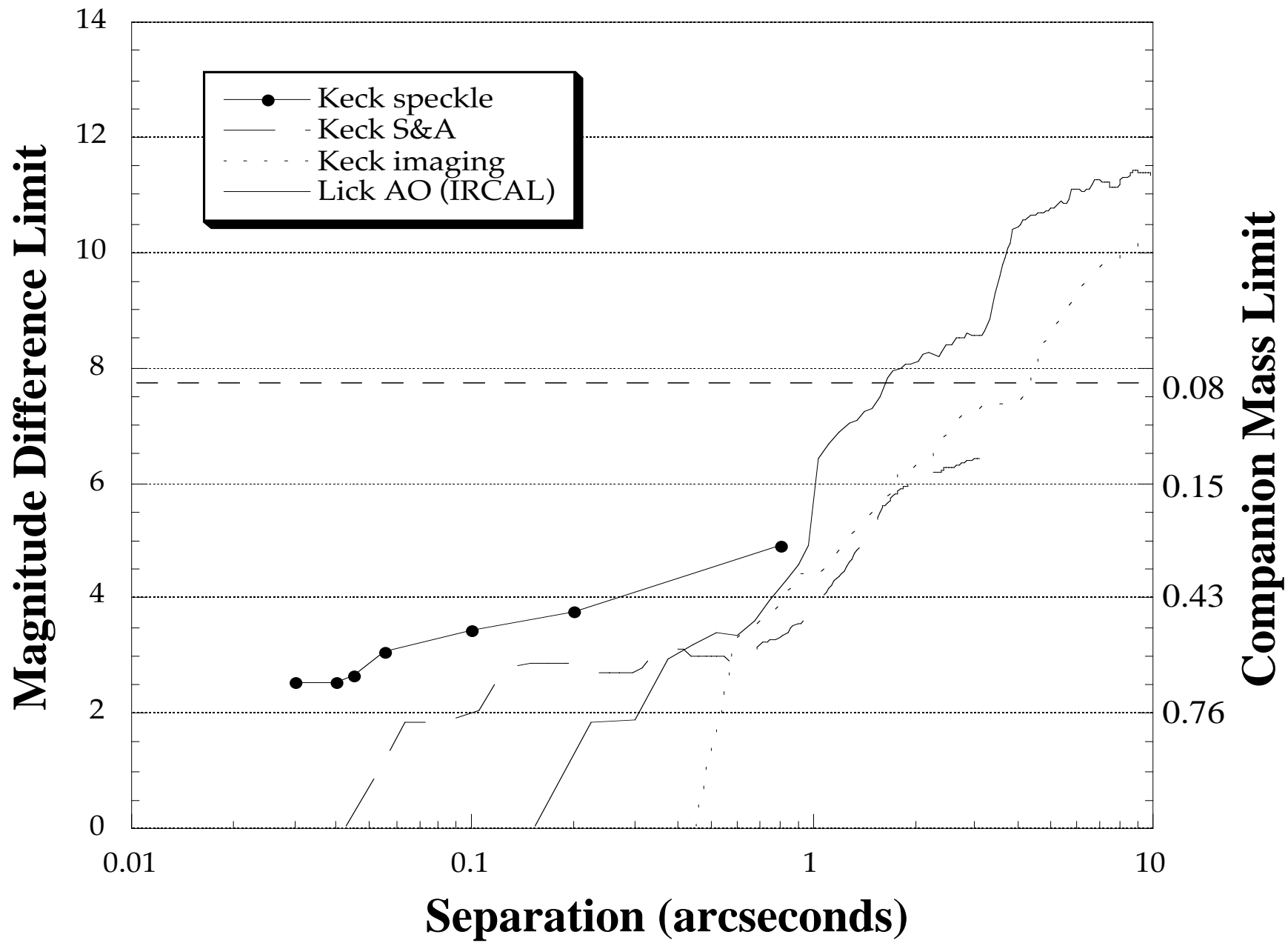

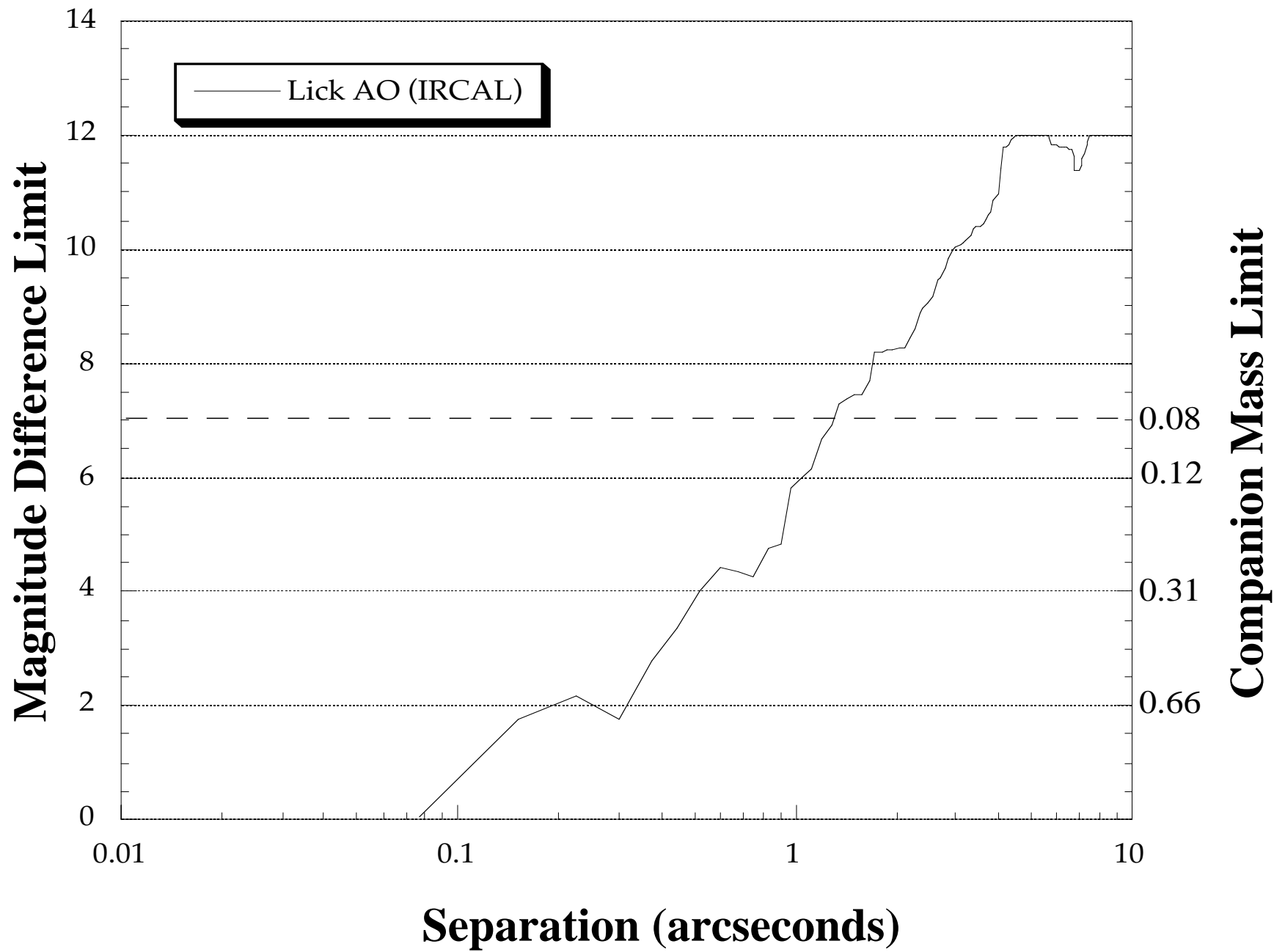

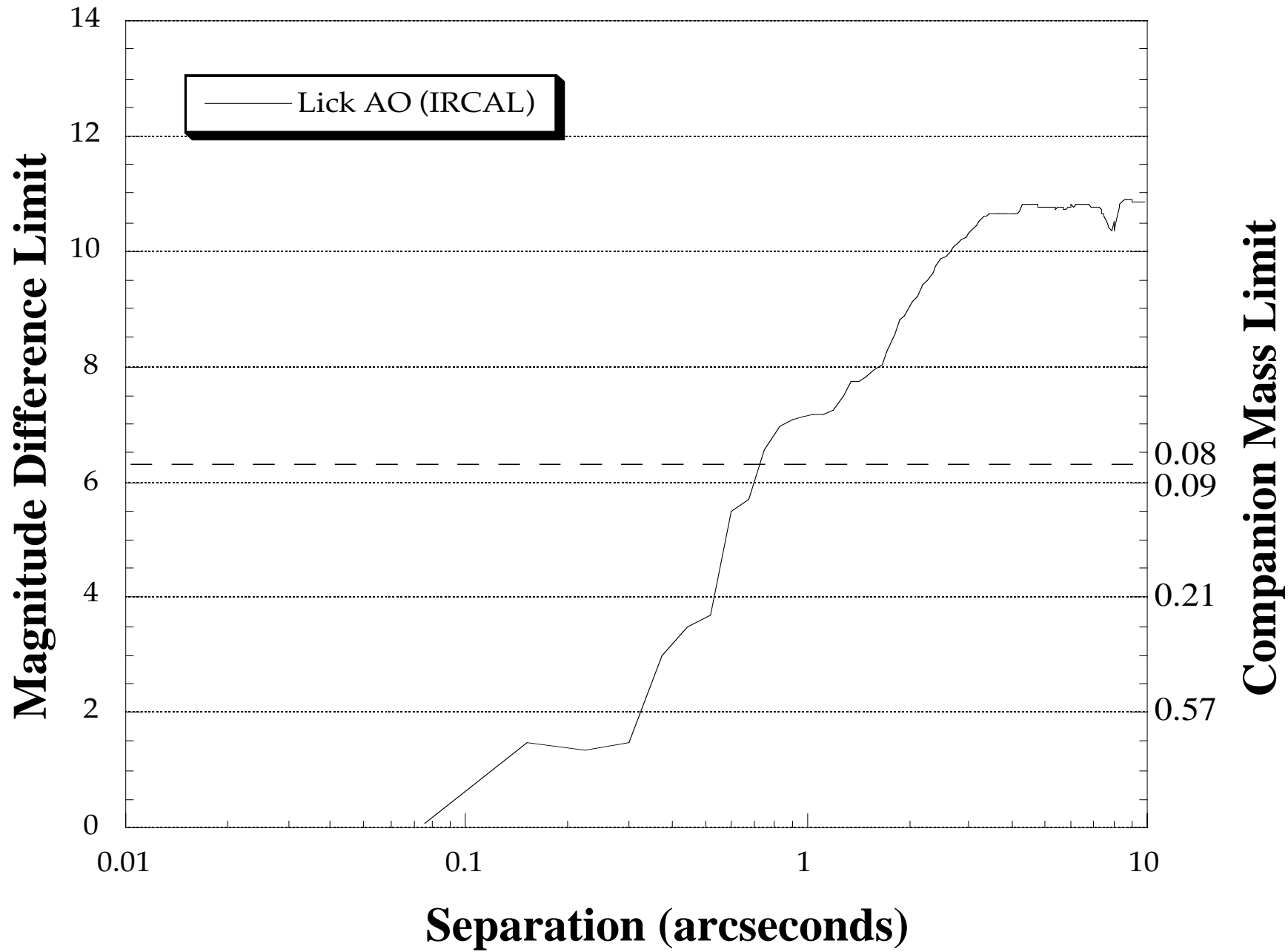

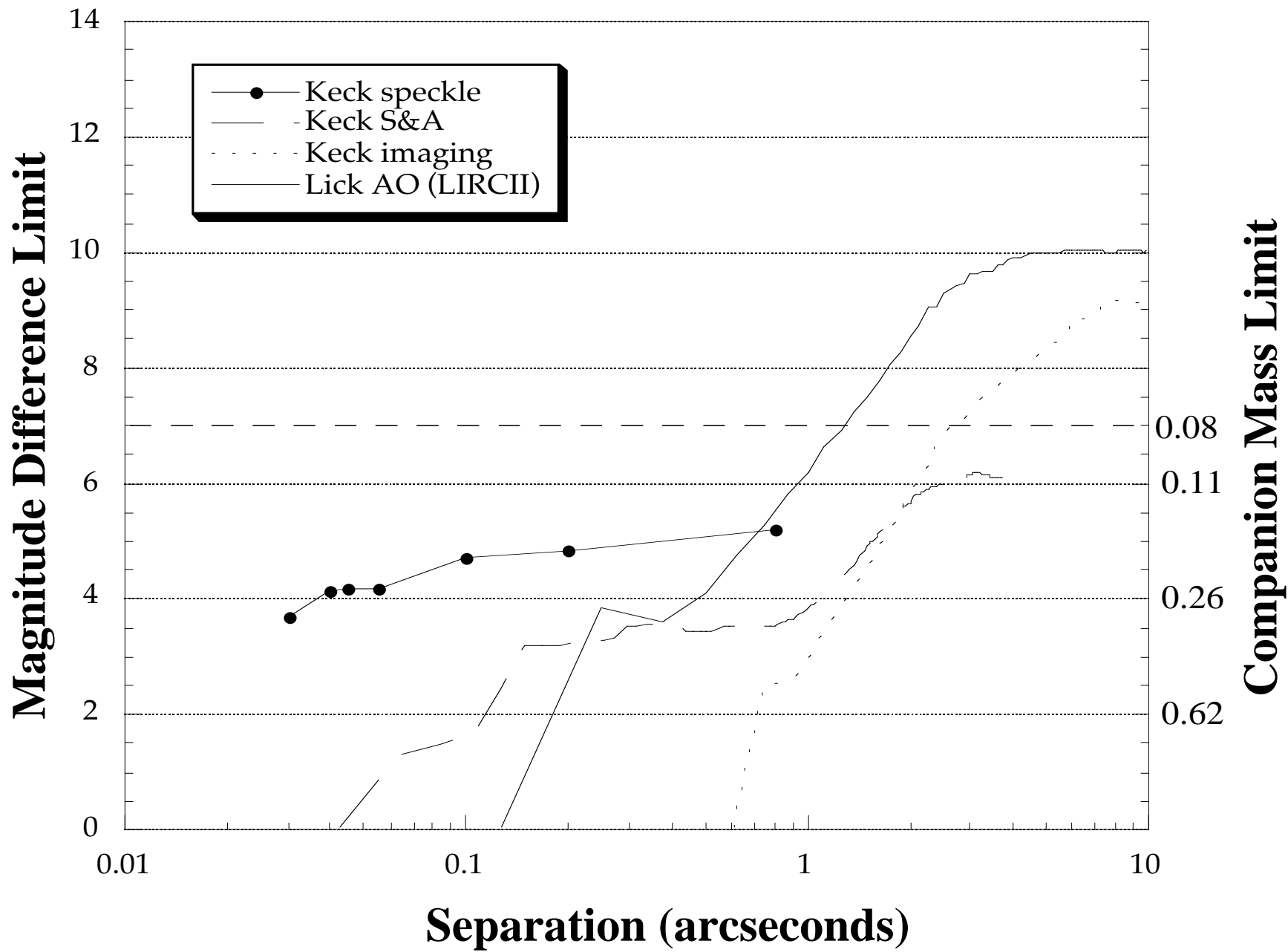
16 Cyg B

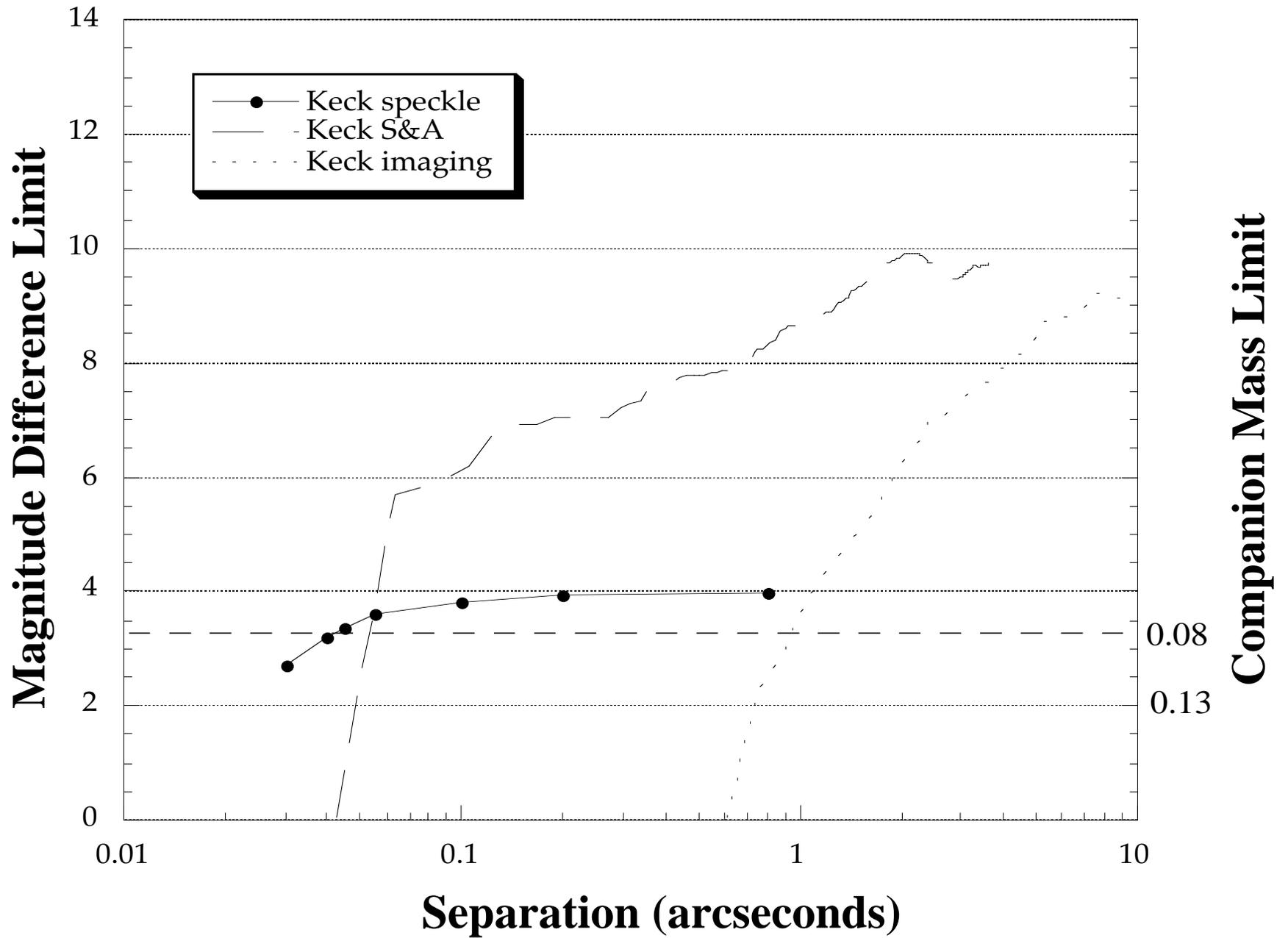

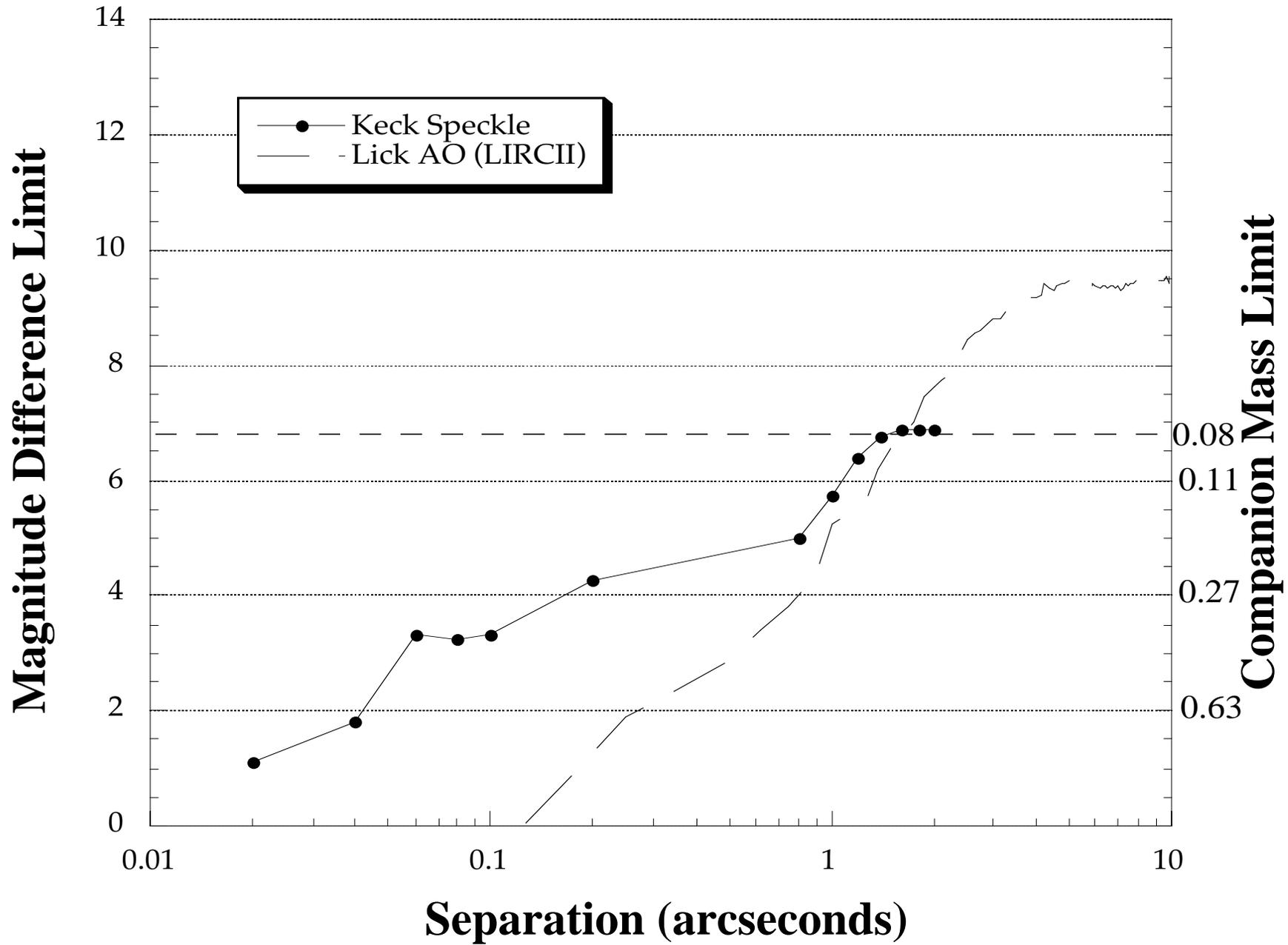

51 Peg

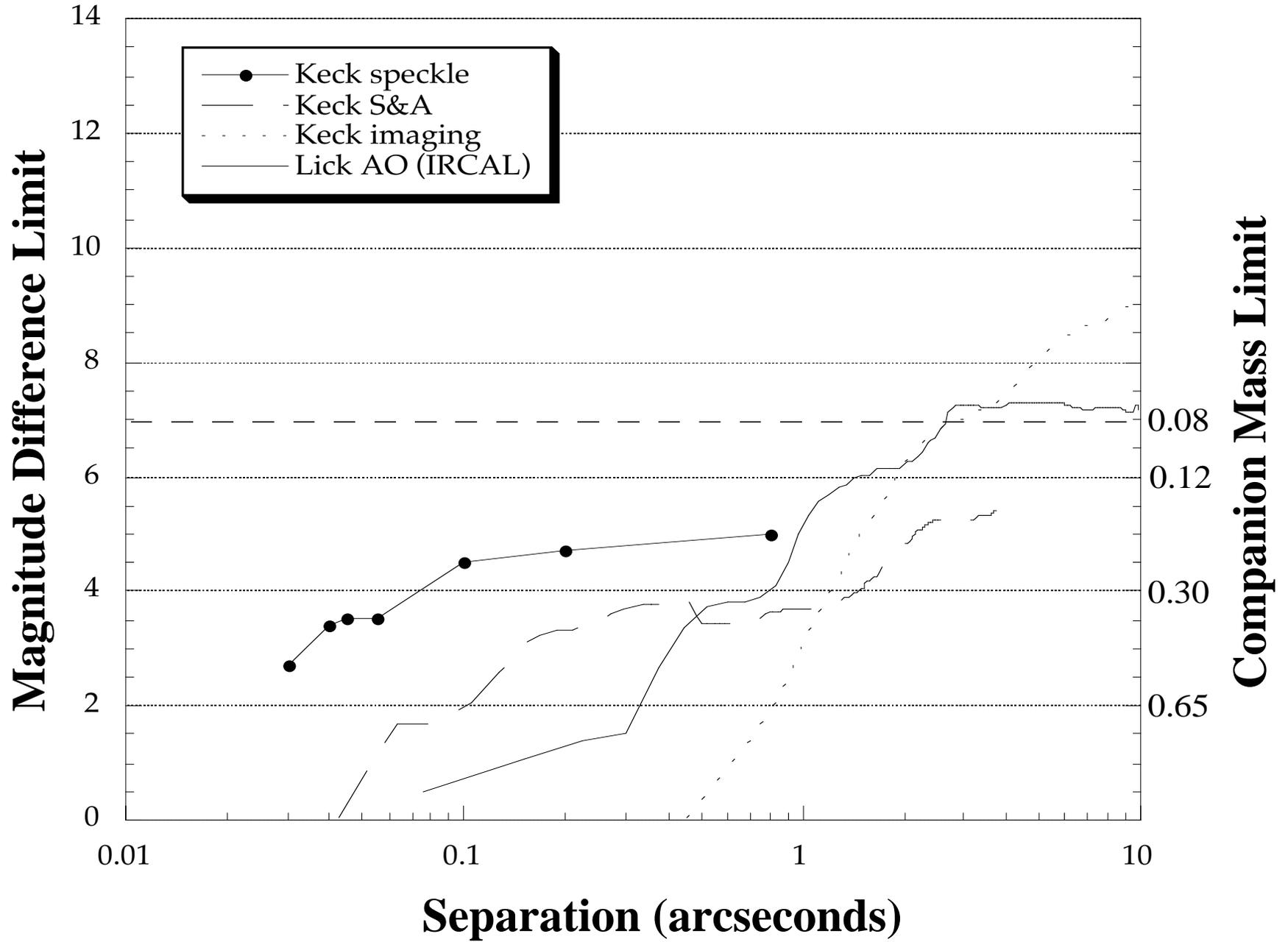

Captions

Figure 1: The median magnitude difference detection limit for unseen companions is plotted as a function of angular separation; the wavelength range is that of the Brγ, K' or $K_s$ filter, depending upon the camera used. The error bars represent one standard deviation of the best limit for each target at a given separation. Individual detection limit plots for each target are given in Appendix A. Only stars observed with Keck speckle imaging are included in the data points with separations less than 0".1. For separations greater than 0".1, the limits are the average of the entire sample. Beginning at separations of 1".5, the median detection limit is $\Delta K=7$ mag which is sensitive enough to detect companions down to the bottom of the main sequence. The detected systems are plotted as crosses.

Figure 2: Images of HD 114762 covering a one and one-half year baseline are shown with the orientation such that North is up and East is to the left. On the left is the central 20" x 20" portion of the OII direct image taken with the NIRC camera on the Keck telescope and on the right is the $K_s$ (2.2 μm) image taken with the IRCAL camera behind the AO system at the Shane telescope. The separation and position angle of the companion remain constant, while a background star would have moved 0".9 to the East in the time between the two images; the orientation of the images places North up and East to the left.

Figure 3: A spectrum of the companion to HD 114762 in the region from 1.15 – 1.35 μm obtained with NIRSPEC behind the Keck AO system in June 2000. The spectrum is normalized by the average of the continuum and the signal-to-noise ratio is greater than 50. The average resolving power is R~2500. Major spectral features are labeled. Only the onset of the strong $H_2O$ (steam) band is indicated. Except for the unusually strong steam band, the spectrum would be consistent with a mid-M star.

Figure A1: The median magnitude difference limit is plotted for each technique employed in the survey – Keck speckle, Keck shift-and-add, Lick AO with the LIRCII and IRCAL cameras, and Keck direct imaging. The symbols marking the different techniques are: a solid line with large dots for Keck speckle, a dashed line for Keck shift-and-add, a dashed-dotted line for Lick AO with LIRCII, a dotted line for Keck imaging, and a solid line for Lick AO with IRCAL. Only the broad or narrowband 2.2 µm data are included and the specific filters used are listed in Table 3.

Figures A2-13: The detection limits as a function of separation are shown for each target and the limits from each technique are plotted individually. Some stars were only observed with a single camera system, but most targets include data from several techniques. The symbols used to indicate the different types of data are: a solid line with large dots for Keck speckle, a dashed line for Keck shift-and-add, a dotted line for Keck imaging, and a solid line for Lick AO. Both the magnitude difference and the corresponding companion mass limit are indicated on the y-axis. The magnitude is either K', $K_s$, or Brγ depending upon which camera was used and the companion mass limits are calculated from the Henry & McCarthy (1993) mass-magnitude relations and the absolute target magnitudes determined from the parameters in Table 1. The horizontal dashed line marks the sensitivity level needed to reach the Hydrogen-burning limit; for most of the sample this limit is achieved between 1 and 2 arcseconds.